\documentclass[referee,manuscript]{gji} 
\usepackage{timet, subfigure, graphicx}
\usepackage{amsmath, txfonts}
\usepackage{multirow,multicol}
\let\bibhang\relax
\usepackage{natbib}

\title[Joint Inversion of Sources and Seismic Waveforms]
  {A Joint Inversion of Sources and Seismic Waveforms for Velocity Distribution: 1-D and 2-D Examples}
\author[Han YU]
{  Han YU $^1$ \\
   $^1$ School of Computer Science, Nanjing University of Posts and
    Telecommunications, Nanjing \emph{210023}, China, \\Email: han.yu@kaust.edu.sa; han.yu@njupt.edu.cn
}

\date{ June 2024 }
\pagerange{\pageref{firstpage}--\pageref{lastpage}}
\volume{}
\pubyear{}

\begin{document}
\label{firstpage}

\maketitle
\begin{summary}
 Waveform inversion is theoretically a powerful tool to reconstruct subsurface structures, but a usually encountered problem is that accurate sources are very rare, causing the computation unstable and divergent. This challenging problem, although sometimes ignored and even imperceptible, can easily create discrepancies in calculated shot gathers, which will then lead to wrong residuals that must be migrated back to the gradients, hence jeopardizing the inverted tomograms. In practice, any shot gather may correspond to its own source even if some of them can be transformed alike after data processing. To resolve this problem, we propose a collocated inversion of sources and early arrival waveforms with the two submodules executing alternatively. Not only can this method produce a decent wavelet that approaches the true source or an equivalent source, but more importantly, it can also invert for credible background velocity models with the optimized sources. Part of the cycle skipping problems can also be mitigated because it avoids the trial and error experiments on various sources. Numerical tests upon a series of different conditions validate the effectiveness of this method. Restrictions on initial sources or starting velocity models will be relaxed with this method, and it can be extended to any other applications for engineering or exploration purposes.
\end{summary}


\section{Introduction}
\indent Full waveform inversion (FWI) is an ambitious but still difficult goal for seismic exploration. Since Tarantola (\citet{tarantola1984inversion}) set up this marvelous target using the Lebesgue measure from the perspective of probability theory, numerous efforts have been made on inverting field datasets, where only a small portion of them can be regarded as successful attempts \cite[]{virieux2009overview}. Immediate causes of this phenomena lie in the implementation, including either the algorithms that are computationally too expensive, or the hardware that is still unsatisfactory in terms of storage or dominant frequency. An important but often ignored problem of FWI comes from the difficulty of estimating accurate sources for the entire flow of data processing and inversion \cite[]{yilmaz2001seismic}. In studying solid earth, this problem has continuously attracted the interests of seismologists in analyzing earthquakes \cite[]{alewine1974application, fukahata2003waveform}, because it is significant to predict aftershocks and prevent secondary disasters. Different from seismic exploration, research of the Earth pays more attention on inverting earthquake signals for their source parameters such as the event magnitude, source tensor, spatial-temporal slip-rate, and possible fault plane orientations \cite[]{yagi2011introduction, zhu2013parametrization}, where theories like Akaike's Bayesian Information Criterion and probability distribution function are often referred to as fundamental tools.
\\\indent In the field of energy exploration and geotechnical investigation, man-made seismic sources are the most frequently used techniques, whose quality has a direct impact on the intermediate images regarded as gradients and the final tomograms \cite[]{bormann2013seismic}. In fact, subsurface structures can be imaged with any reasonable input sources, because the back-projected data largely determine the migration resolution. A typical example is the time-reversal imaging method by Gajewski $et~al.$ (\citeyear{gajewski2005reverse}) and Fink (\citeyear{fink2006time}), who applied migration to locating passive seismic sources \cite[]{baysal1983reverse}. Later on, various improvements based on this method have been made either to achieve attenuation compensation \cite[]{zhu2014time} and high resolution \cite[]{liu2021sparsity}, or to extend it to elastic waves \cite[]{artman2010source}, or to update the imaging condition by zero-lag correlation \cite[]{sun2016full}. Even if we conduct least-squares migration on a mediocre velocity model, incorrect sources may still lead to convergent images in that the product of the linear forward modeling operator and its adjoint counterpart is always positive semi-definite \cite[]{dai2011least}.
\\\indent However, for inversions of surface or vertical seismic profiles, the credibility of inverted tomograms highly relies on the quality of sources rather than their locations. In FWI, one of the most frequently encountered problems stems from an inaccurate background velocity distribution, indicating the high nonlinearity of model space with respect to the data space \cite[]{schuster2017seismic}. Not surprisingly, inversion ``crimes'' are usually made by either creating unrealistically good starting models, or sifting out ultra-low frequency data through deconvolution, or enabling neural networks to invert similar datasets ever trained as input in recent studies \cite[]{sun2021physics}. As a matter of fact, the sources, as well as the adjoint sources, are closely related to all the aspects mentioned above. Inappropriate sources, even with small perturbations, can eventually cause the famous ``cycle skipping'' problem, hence making the optimization process trapped in local minima. In this situation, successful data fitting of some shot gathers rarely happens because incorrect sources may destroy low-wavenumber parts of the velocity models.
\\\indent On the one hand, from the perspective of computation, a small perturbation in the source wavelet can introduce considerable discrepancies into the seismograms, causing noticeable errors in the updated tomograms. Moreover, it will reversely come back and mislead the source estimation because inaccuracies of wavelets can be strengthened iteratively. Conventionally, first arrivals extracted from near-offset traces \cite[]{sheng2006early} are usually treated as sources to deal with this problem; nevertheless, some diehard gathers are impossible to be matched well, preventing the velocity from approaching the right distribution. On the other hand, from the data acquisition aspect, every common shot gather (CSG) corresponds to its own source although they can be more or less processed to maximize their overlapped frequency spectra. Even worse, matching filters may inject new distortion into the raw data, leading to unexpected effects on the tomograms. Very near-offset traces are sometimes dropped in FWI \cite[]{lamuraglia2023application} owing to their strong energy that may dominate the data matching procedure or their severely distorted amplitudes that are not in accordance with the propagating rules by the wave equation.
\\\indent To partly solve these problems, van den Berg $et~ al.$ have developed and improved the contrast source inversion method to invert the sources and media simultaneously \cite[]{van1997contrast, abubakar2008finite} for imaging electromagnetic inverse scatters. Different from seismic imaging, their high frequency sources are located on a curve outside of and surrounding a simply connected and bounded domain. Here, the measurement of the scatter field is made with a homogenous background model. Another assumption that the scatterers are illuminated successively by a number of incident waves of one single frequency is usually not satisfied in seismic imaging. Symes $et~ al.$ have constructed an extended source function to fit the recorded data by solving a source matching subproblem \cite[]{huang2018source, symes2020full}, and estimated the errors in transmission inverse problems \cite[]{symes2022error}. However, the nonphysical space-time extended source model depends on spatial and time variables, which cost large computer memory for storing these functions. Moreover, the annihilator, as a regularization term to describe source-receiver-dependent signatures, can add a few more uncertainties and expenses to computation \cite[]{van2013mitigating}, since it requires storing all of the time slices of the Green's function, which makes the implementation possible in the frequency domain only. Other methods to get around this problem are more or less related to modifying the objective functionals with techniques like convolution or deconvolution \cite[]{yu2017application, zhong2019source} so as to increase the data fitting probabilities, although they are still not computationally economic \cite[]{lee2003source}.
\\\indent It is thus proposed in this study to estimate the accurate or equivalent sources as well as inverting the background velocity model alternatively. With the help of early arrival waveform inversion (EWI), we follow the work done my Wang $et~ al.$ (\citeyear{wang2009simultaneous}) and Coats $et~al.$ (\citeyear{sun2014source}), but using the early arrivals instead of full waveforms. One reason is that FWI may not back-propagate the received energy to the source position owing to acquisition properties, whereas EWI can easily mitigate this problem. Another reason is that early arriving signals are more linear with respect to the velocity perturbations. Thirdly, EWI focuses on the background velocity model rather than its high wavenumber parts. Therefore, it will in turn act positively on the source inversion, which can pull the inverted sources off the local minima. A complete series of numerical tests on several one-dimension and two-dimension models with generalized initial velocity models and sources will show the effectiveness and robustness of this approach.
\\\indent This article is composed of four parts. After the introductory section, we will describe the detailed theory of source inversion accompanied by EWI in Section 2. Also, a methodological framework of the proposed collocated alternative inversion of sources and waveforms will be presented for the following implementations. Section 3 presents a series of complete numerical tests on both the one-dimensional and the two-dimensional cases with various sources to illustrate the effectiveness and generality of the algorithms. Section 4 is the conclusive part, where possible applications and extensions of this study will also be discussed.


\section{Theory and Method}
Given some initial velocity models or sources, we will present the framework of the collocated inversion of sources and early arrival waveforms alternatively. These two modules are relatively independent, but they may promote or undermine each other's function intrinsically with certain numbers of executions assigned to them. We first present the theory of source inversion based on the acoustic wave equation, which literally unveils the implementation techniques. Secondly, we briefly review the methodology of EWI, followed by the workflow of the proposed collocated inversion of the two modules in the final subsection.



\subsection{The Source Inversion Method}
\label{sec2A}
To derive the Fr\'echet derivative of the pressure field $p$ with respect to the source perturbation, we take the acoustic wave equation as an example to present the source inversion scheme:
\begin{eqnarray}
    \frac{1}{c^2({\bf{x}})}\frac{\partial^2 p({\bf{x}},t;{\bf{x}}_s)}{\partial t^2}-\rho({\bf{x}})\nabla\cdot({\frac{1}{\rho({\bf{x}})}}\nabla p({\bf{x}},t;{\bf{x}}_s))=src(t;{\bf{x}}_s),
    \label{eqn_acou}
\end{eqnarray}
\noindent where $p({\bf{x}},t|{\bf{x}}_s)$ represents the traces recorded at position ${\bf{x}}$, time $t$ with a point source $src(t;{\bf{x}}_s)$ excited at ${\bf{x}}_s$, which propagates through the media with its velocity distribution $c({\bf{x}})$ and density vector $\rho(\bf{x})$, respectively. A classical solution to the above equation (\ref{eqn_acou}) with a delta function source exciting at time $t'$ and position ${\bf{x'}}$ is the Green's function $g$ that obeys the following equation:
\begin{eqnarray}
    \nonumber\frac{1}{c^2({\bf{x}})}\frac{\partial^2 g({\bf{x}},t;{\bf{x'}},t')}{\partial t^2}-\rho({\bf{x}})\nabla\cdot({\frac{1}{\rho({\bf{x}})}}\nabla g({\bf{x}},t;{\bf{x'}},t'))=
    \\\delta({\bf{x-x'}})\delta(t-t'),
    \label{eqn_deltaSln}
\end{eqnarray}
\noindent where the following conditions are satisfied
\begin{eqnarray}
     \nonumber g({\bf{x}},t;{\bf{x'}},t') = 0,~ and~ \dot{g}({\bf{x}},t;{\bf{x'}},t') = 0, ~for~ t\leq t'.
\end{eqnarray}
\indent Suppose a little perturbation of the source $\delta_{src}(t;{\bf{x}}_s)$ can cause a corresponding variation $\delta_1{p}$ in the received pressure profiles, and the equation can be rearranged as
\begin{eqnarray}
    \nonumber \frac{1}{c^2({\bf{x}})}\frac{\partial^2 (p+\delta_1{p})}{\partial t^2}-\rho({\bf{x}})\nabla\cdot({\frac{1}{\rho({\bf{x}})}}\nabla (p+\delta_1{p}))=
    \\src(t;{\bf{x}}_s)+\delta_{src}(t;{\bf{x}}_s).
    \label{eqn_acou_srcPerturb}
\end{eqnarray}
\noindent Subtracting (\ref{eqn_acou}) from (\ref{eqn_acou_srcPerturb}) gives a new equation
\begin{eqnarray}
    \frac{1}{c^2({\bf{x}})}\frac{\partial^2 \delta_1{p}}{\partial t^2}-\rho({\bf{x}})\nabla\cdot({\frac{1}{\rho({\bf{x}})}}\nabla \delta_1{p})=\delta_{src}(t;{\bf{x}}_s) + O(\delta_{src}^2),
    \label{eqn_acou_pressurePerturb}
\end{eqnarray}
\noindent where the high order term of source perturbation can be neglected with out loss of generality. So the solution can be rewritten in terms of the Green's function by convolution:
\begin{eqnarray}
    \delta_1{p}({\bf{x}}_r,t;{\bf{x}}_s) =
    \int_{v}g({\bf{x}}_r,t;{\bf{x'}},0)\ast\delta_{src}(t;{\bf{x}}_s) ~dv({\bf{x'}}).
    \label{eqn_slnPressurePerturb}
\end{eqnarray}
\indent Now we assume that the delta function can be rewritten as a combination of the temporal and the spatial terms by multiplication
\begin{eqnarray}
    \delta_{src}(t;{\bf{x}}_s)=\Delta src(t)\delta_{src}({\bf{x}}-{\bf{x}}_s),
    \label{eqn_srcCombTimeSpace}
\end{eqnarray}
\\\noindent because it is reasonable to consider the occurrence of source perturbation only at the source location. Consequently, the perturbed wavefield becomes
\begin{eqnarray}
    \delta_1{p({\bf{x}}_r,t;{\bf{x}}_s)}=g({\bf{x}}_r,t;{\bf{x}}_s,0)\ast\Delta src(t).
    \label{eqn_pressurePerturb}
\end{eqnarray}
\noindent Dividing both sides of equation (\ref{eqn_pressurePerturb}) by $\Delta src(t)$ shall we estimate the derivative of the calculated pressure field with respect to the source perturbation by
\begin{eqnarray}
    \nonumber\frac{\partial{p_{calc}({\bf{x}}_r,t;{\bf{x}}_s)}}{\partial{src}} &= g({\bf{x}}_r,t;{\bf{x}}_s,0) \ast\delta_{src}(t'-t) \\ &= g({\bf{x}}_s,t;{\bf{x}}_r,0)\ast\delta_{src}(t'-t),
    \label{eqn_derivativePressureSrc}
\end{eqnarray}
\noindent where the last equality is by the Reciprocity theorem.
\\\indent Now if the misfit functional is defined by
\begin{eqnarray}
    J=\frac{1}{2}\underset{s,r}{\sum}\int{(\Delta{p({\bf{x}}_r,t;{\bf{x}}_s,0)})^2~dt},
    \label{eqn_misfit}
\end{eqnarray}
\noindent the discrepancies between the calculated and the observed early arrival waveforms can be expressed as
\begin{eqnarray}
    \nonumber\Delta{p} =& (p_{calc}({\bf{x}}_r,t;{\bf{x}}_s,0) - p_{obs}({\bf{x}}_r,t;{\bf{x}}_s,0)) W({\bf{x}}_r,t;{\bf{x}}_s,0)
    \\ = & p^{win}_{calc}({\bf{x}}_r,t;{\bf{x}}_s,0) - p^{win}_{obs}({\bf{x}}_r,t;{\bf{x}}_s,0),
    \label{eqn_deltaEarr}
\end{eqnarray}
\noindent where $W$ is a time window that constraints meaningful signals. The objective function can be rearranged in terms of the source indices $s$ as
\begin{eqnarray}
    J = \underset{s}{\sum} J_s,
    \label{eqn_misfitSrcs}
\end{eqnarray}
\noindent where $J_s$ denotes
\begin{eqnarray}
    J_s = \frac{1}{2}\underset{r}{\sum}\int(p^{win}_{calc}({\bf{x}}_r,t;{\bf{x}}_s,0) - p^{win}_{obs}({\bf{x}}_r,t;{\bf{x}}_s,0))^2~ dt.
    \label{eqn_misfitOneSrc}
\end{eqnarray}
\indent If we only update the $src$ located at ${\bf{x}}_s$ by freezing the velocity distribution, then the gradient, for each source, can be expressed by
\begin{eqnarray}
    \nonumber d^{src} =& \frac{\partial J_s}{\partial src} = (p^{win}_{calc} - p^{win}_{obs})\frac{\partial p^{win}_{calc}}{\partial src}
    \\ \approx & \underset{r}{\sum}g({\bf{x}}_s,t;{\bf{x}}_r,0)\ast(p^{win}_{calc} - p^{win}_{obs}).
    \label{eqn_gradSrc}
\end{eqnarray}
\\\indent Equation (\ref{eqn_gradSrc}) essentially indicates that back-propagating the residual signals from the receivers' sides enables us to optimize the $src$ for each CSG. Therefore every source can be updated by
\begin{eqnarray}
    src_{k+1} = src_{k}(t) + \alpha_kd^{src}_k(t),
    \label{eqn_srcUpdate}
\end{eqnarray}
\noindent where $k$ is the iteration number and $\alpha$ denotes the step length, which is determined by the steepest descent algorithm \cite{nocedal2006}.


\subsection{Early Arrival Waveform Inversion Revisited}
\label{sec2B}
\indent To achieve the goals above, it is not possible to make a successful source update without an accurate background velocity model. Therefore, we will briefly review the process of EWI in this subsection. Based on the misfit function (\ref{eqn_misfit}) defined in last subsection, the following velocity update formula can be derived:
\begin{eqnarray}
    \frac{\partial J}{\partial c({\bf{x}})} = \frac{1}{c^3({\bf{x}})}\underset{s}{\sum}\int{\dot{p}_{fwd}({\bf{x}},t;{\bf{x}}_s,0)\dot{p}_{back}({\bf{x}},t;{\bf{x}}_s,0)~dt},
    \label{eqn_velFrechet}
\end{eqnarray}
\noindent where $\dot{p}$ represents the time derivative of $p$, and $p_{fwd}$ is the forward wavefields whereas $p_{back}$ means the back-projected waveform residuals produced by
\begin{eqnarray}
    p_{back}({\bf{x}},t;{\bf{x}}_s,0) = \int g({\bf{x}},-t;{\bf{x'}},0)\ast\delta p_{back}({\bf{x'}},t;{\bf{x}}_s)~ dx'.
    \label{eqn_backProjResWave}
\end{eqnarray}
\noindent Here, the $p_{back}$ is substantially
\begin{eqnarray}
    \delta p_{back}({\bf{x'}},t;{\bf{x}}_s) = \underset{r}{\sum}\delta({\bf{x'}}-{\bf{x}}_r)(p^{win}_{calc}({\bf{x}}_r,t;{\bf{x}}_s,0) - p^{win}_{obs}({\bf{x}}_r,t;{\bf{x}}_s,0))
    \label{eqn_resWave}
\end{eqnarray}
\\\indent Note that the velocity distribution can be iteratively updated along the conjugate directions defined by
\begin{eqnarray}
    d_k = -P_kgrad_k+\beta_kd_{k-1},
    \label{eqn_conjGrad}
\end{eqnarray}
\noindent where
\begin{eqnarray}
    \beta_k = \frac{grad^T_k(P_k - P_{k-1}grad_{k-1})}{grad^T_{k-1}P_{k-1}grad_{k-1}}.
    \label{eqn_conjStepLen}
\end{eqnarray}
\noindent Here, subscripts $k=1,2,..., k_{max}$ represent the iteration numbers of updating velocity models, and $P$ is the conventional geometrical-spreading preconditioner. The term $grad$ represents the migration image $grad({\bf{x}})$, and for the first iteration, we have $d_0 = -grad_0$. The parameter $\beta_k$ is computer by the Polak-Ribiere formula \cite{nocedal2006}. Therefore, the new velocity distribution is searched by
\begin{eqnarray}
    c_{k+1}({\bf{x}}) = c_{k}({\bf{x}}) +\lambda_kd_k({\bf{x}}),
    \label{eqn_velUpdateDirect}
\end{eqnarray}
\noindent where $\lambda_k$ is the step length which can be determined by aquadratic line-search method \cite{nocedal2006}, and $d_k({\bf{x}})$ is the component of the direction vector indexed by $k$. To compute the gradient direction for each iteration reduces to the reverse time migration operation, and additional forward modelings are required for the line search.


\subsection{The Collocated Inversion Strategy}
\label{sec2C}
\begin{figure*}
\centering
    \subfigure{\includegraphics[width=3.5in]{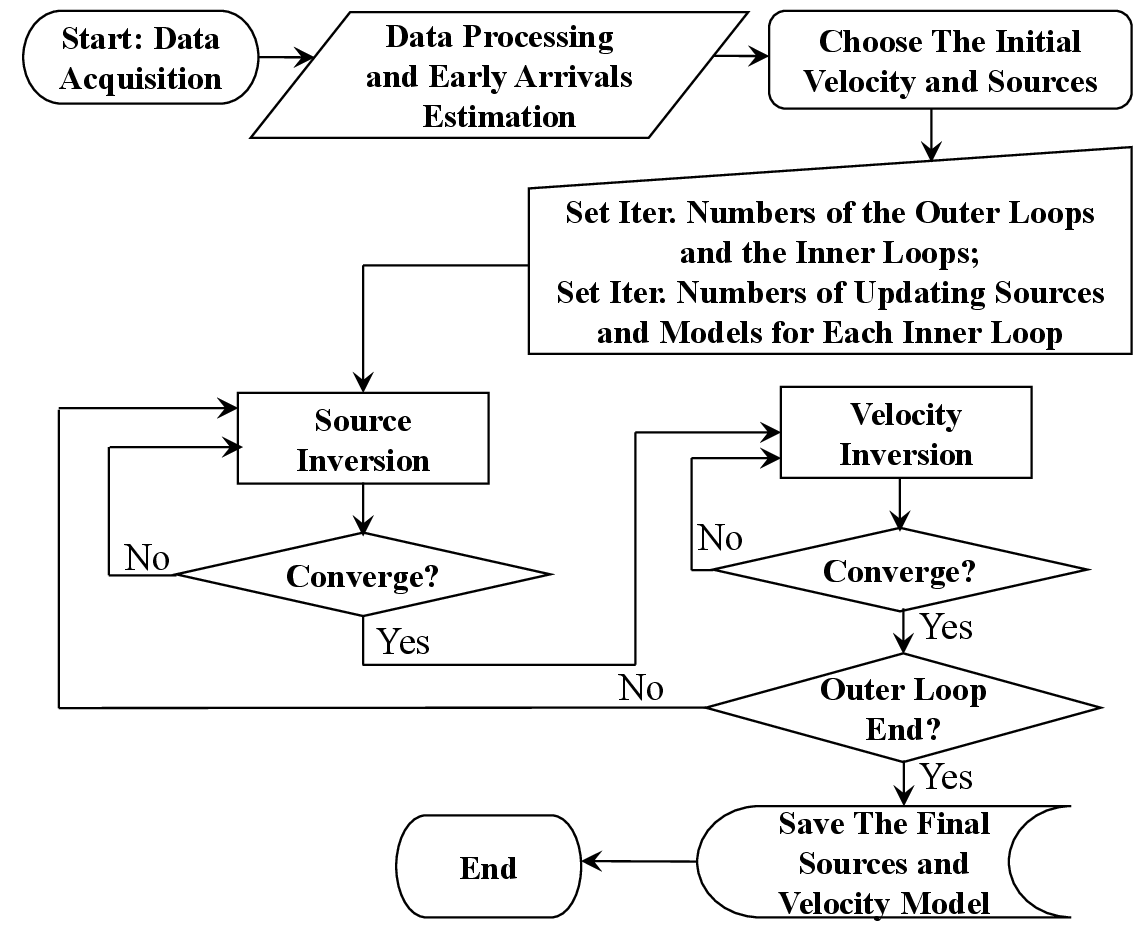}}
    \caption{Flowchart for the collocated inversion of sources and early arrival waveforms alternatively. }
\label{fig1_workFlow}
\end{figure*}

\indent The two submodules, namely the source inversion and EWI, should cooperate with each other, because the initial sources cannot head towards correct directions without the support of a decent background velocity and vice versa. The framework of this collocated inversion is presented in Fig. \ref{fig1_workFlow}, which indicates the alternating participation of the two modules in the entire loop. If the parameters, such as the number of iterations of source inversion and EWI, are reasonably setup, both the sources and the velocity model can take turns to converge iteratively such that these two profiles can approximate their ground truths.
\\\indent In general, the preprossing steps after data acquisition are similar routines, and only early arrivals are our target waveforms in this work. This method does not exert strict constraints on the accuracies of the input initial sources and velocity model. However, we need to set up reasonable numbers of iterations for the outer and the inner loops at first. Moreover, within every inner loop, independent rounds of iterations are assigned to the source inversion and EWI.
\\\indent In our empirical tests, 1:2 or 1:3 is an effective proportion for the two types of inversions. Namely, if the sources are updated once, then the velocity model needs to be updated twice or three times. In practice, it is natural that every source should be unique in some sense so that all of them need to be inverted separately even if they can be filtered to be similar.


\section{Numerical Experiments}
\indent To validate the proposed method, two series of numerical tests are taken for illustrating the effectiveness of the workflow as well as the associated techniques. The first subsection shows numerical results on several one-dimensional cases with different initial models and sources. The second subsection shows numerical results on a classical two-dimensional SEG/EAGE Overthrust model with various input sources and velocity distributions.


\subsection{One-Dimensional Examples}
\begin{figure*}
\centering
\includegraphics[width=3in]{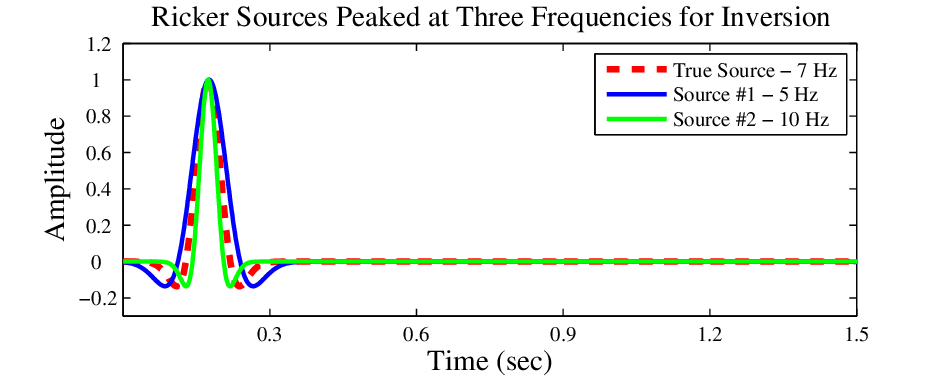}
    \caption{The Ricker source wavelets peaked at three different frequencies. }
\label{fig2_1D_srcsAlign}
\end{figure*}

\begin{figure*}
\centering
\includegraphics[width=2in]{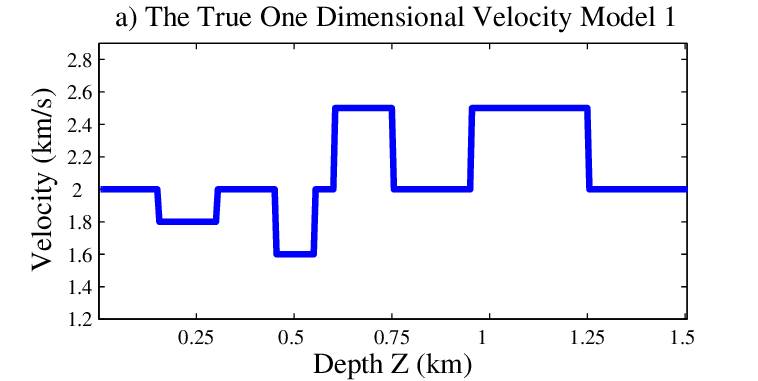}
\includegraphics[width=2in]{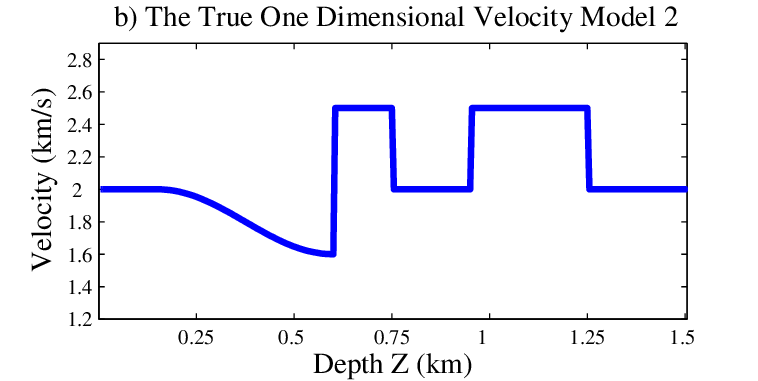}
    \caption{Two true velocity models, (a) one with a layered near surface velocity distribution, and (b) the other with a gradient near surface velocity distribution. }
\label{fig3_1DTrueModels}
\end{figure*}

\begin{figure*}
\centering
\includegraphics[width=4.5in]{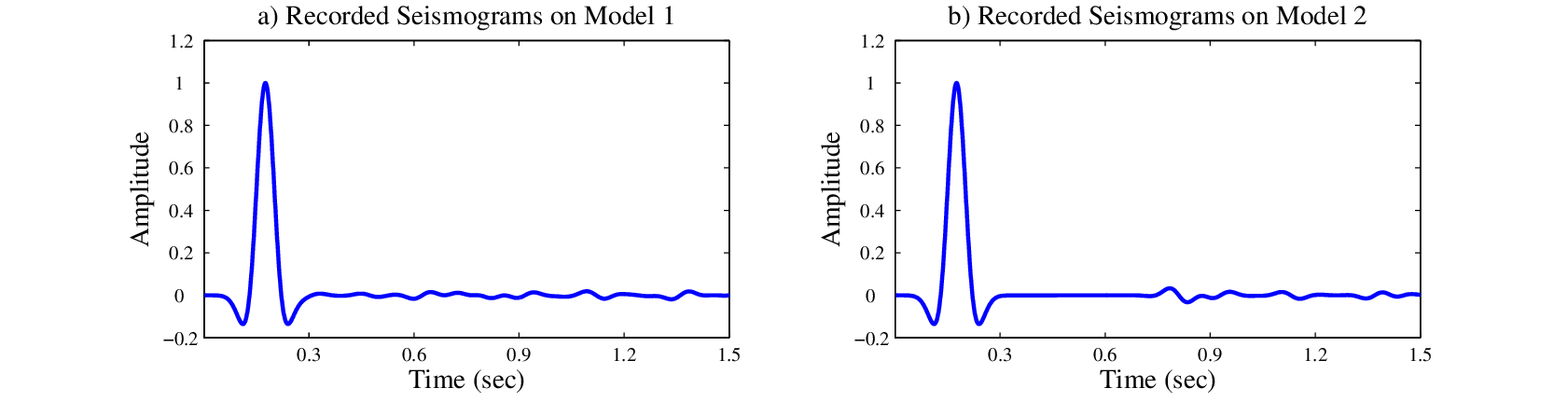}
    \caption{Two seismograms (a) and (b) recorded with the true source propagating on Figs. \ref{fig3_1DTrueModels}(a) and \ref{fig3_1DTrueModels}(b), respectively.}
\label{fig4_1DSeismograms}
\end{figure*}

\begin{figure*}
\centering
\includegraphics[width=2in]{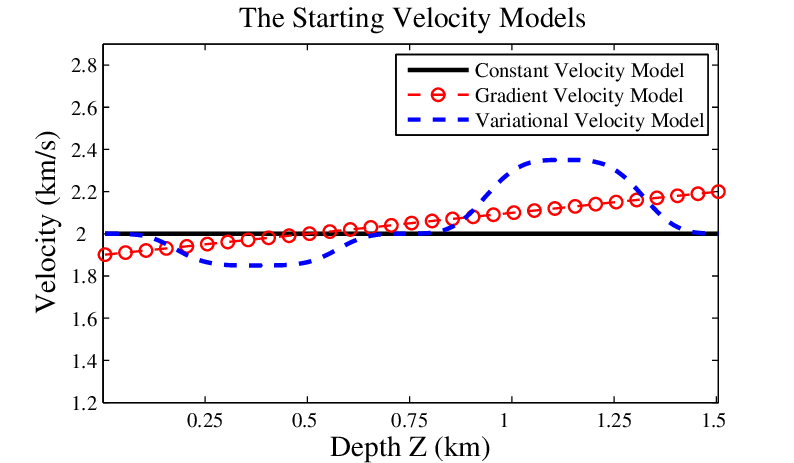}
    \caption{Three one-dimensional starting velocity models.}
\label{fig5_1DInitialVel}
\end{figure*}

\begin{figure*}
\centering
\includegraphics[width=4.5in]{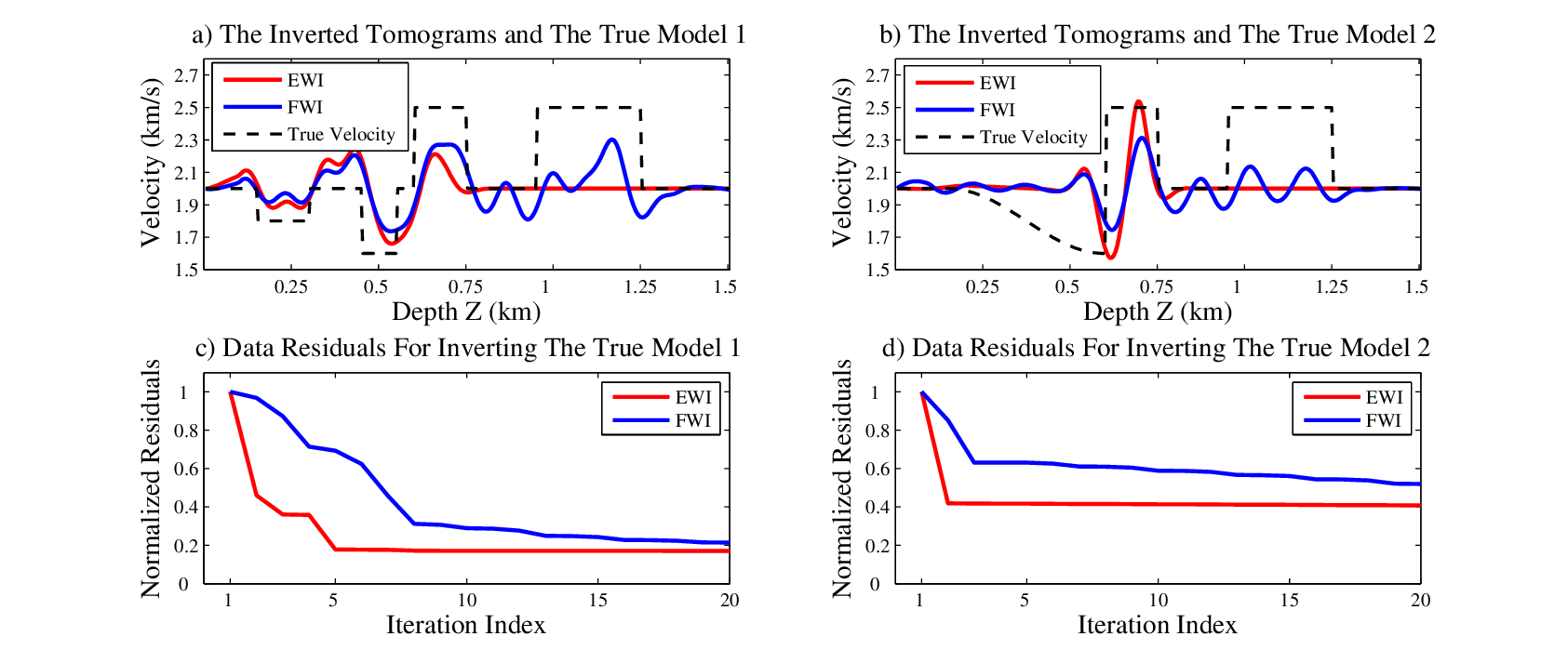}
    \caption{Velocity models (a) and (b) inverted from the constant velocity model with the true source wavelet using EWI and FWI, and their associated data residuals (c) and (d).}
\label{fig6_1DInvConstModels_trueSrc}
\end{figure*}

\begin{figure*}
\centering
\includegraphics[width=4.5in]{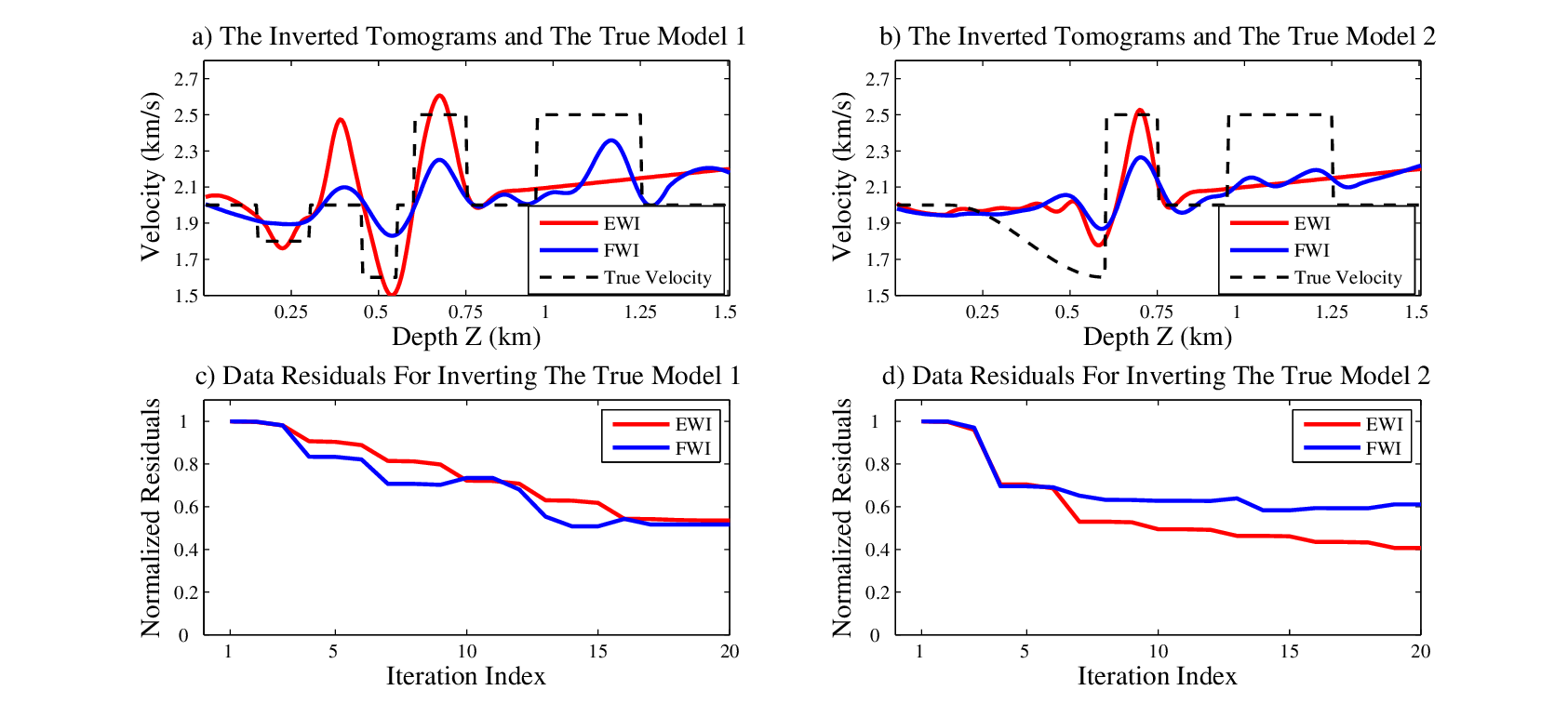}
    \caption{Velocity models (a) and (b) inverted from the gradient velocity model with the true source wavelet using EWI and FWI, and their associated data residuals (c) and (d).}
\label{fig7_1DInvGradModels_trueSrc}
\end{figure*}

\begin{figure*}
\centering
\includegraphics[width=4.5in]{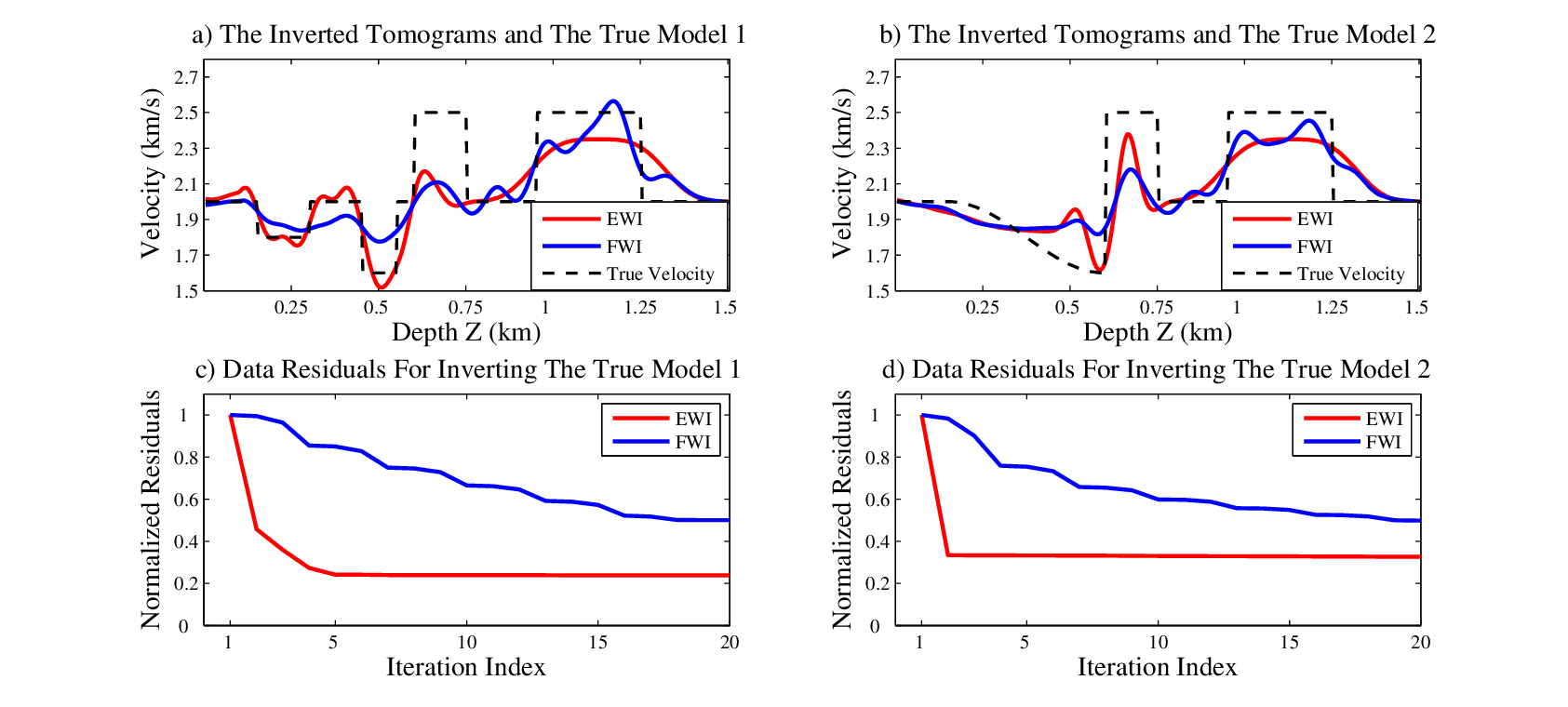}
    \caption{Velocity models (a) and (b) inverted from the variational velocity model with the true source wavelet using EWI and FWI, and their associated data residuals (c) and (d).}
\label{fig8_1DInvVarModels_trueSrc}
\end{figure*}

\begin{figure*}
\centering
\includegraphics[width=4.5in]{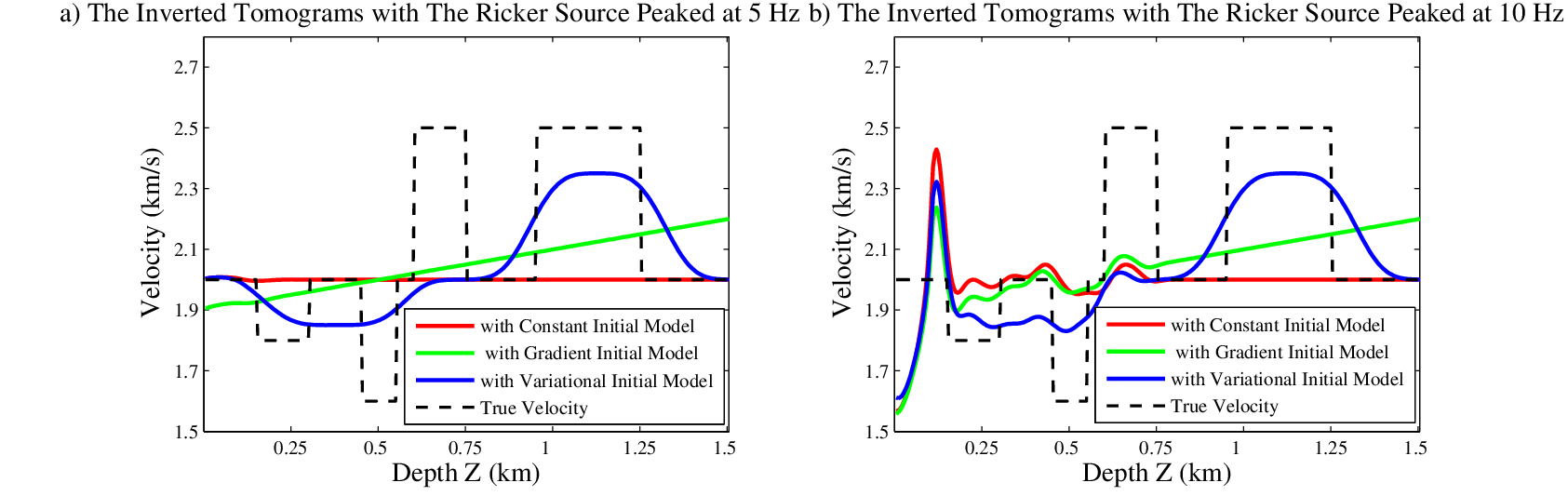}
    \caption{Tomograms inverted from the three starting velocity models in Fig. \ref{fig5_1DInitialVel} by two sources peaked at (a) 5 Hz and (b) 10 Hz, respectively. }
\label{fig9_1DInv3Models_wrongSrcs}
\end{figure*}

\begin{figure*}
\centering
\includegraphics[width=4.5in]{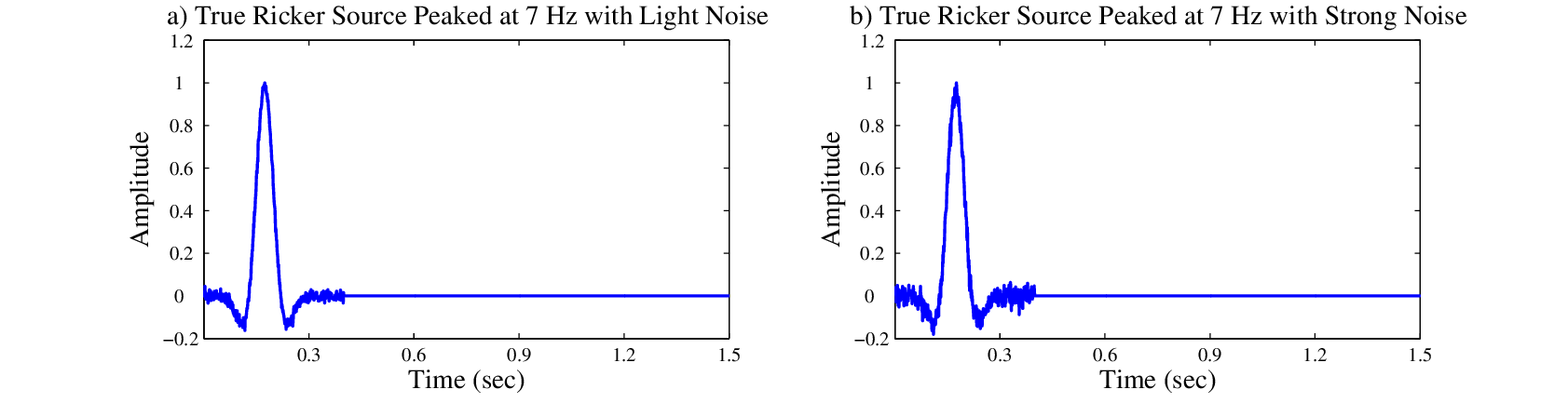}
    \caption{Two Ricker sources peaked at 7 Hz, respectively with (a) light and (b) strong noise.}
\label{fig10_1DNoiseSrcs}
\end{figure*}

\begin{figure*}
\centering
\includegraphics[width=4.5in]{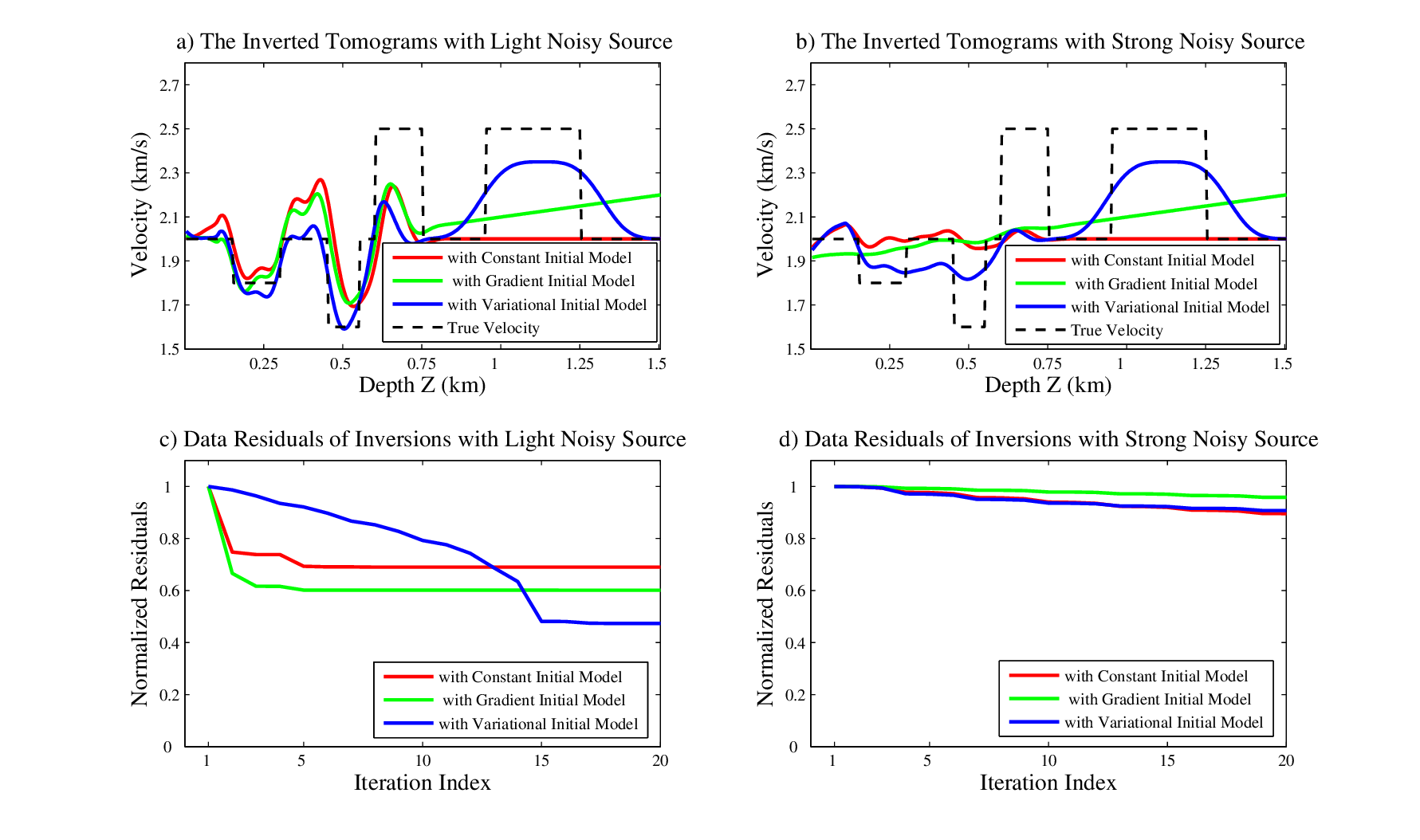}
    \caption{ Tomograms (a) and (b) inverted with noisy sources in Figs. \ref{fig10_1DNoiseSrcs}(a) and \ref{fig10_1DNoiseSrcs}(b), and their associated data residual curves (c) and (d). }
\label{fig11_1DInv3Models_noiseSrcs}
\end{figure*}

\begin{figure*}
\centering
\includegraphics[width=6.5in]{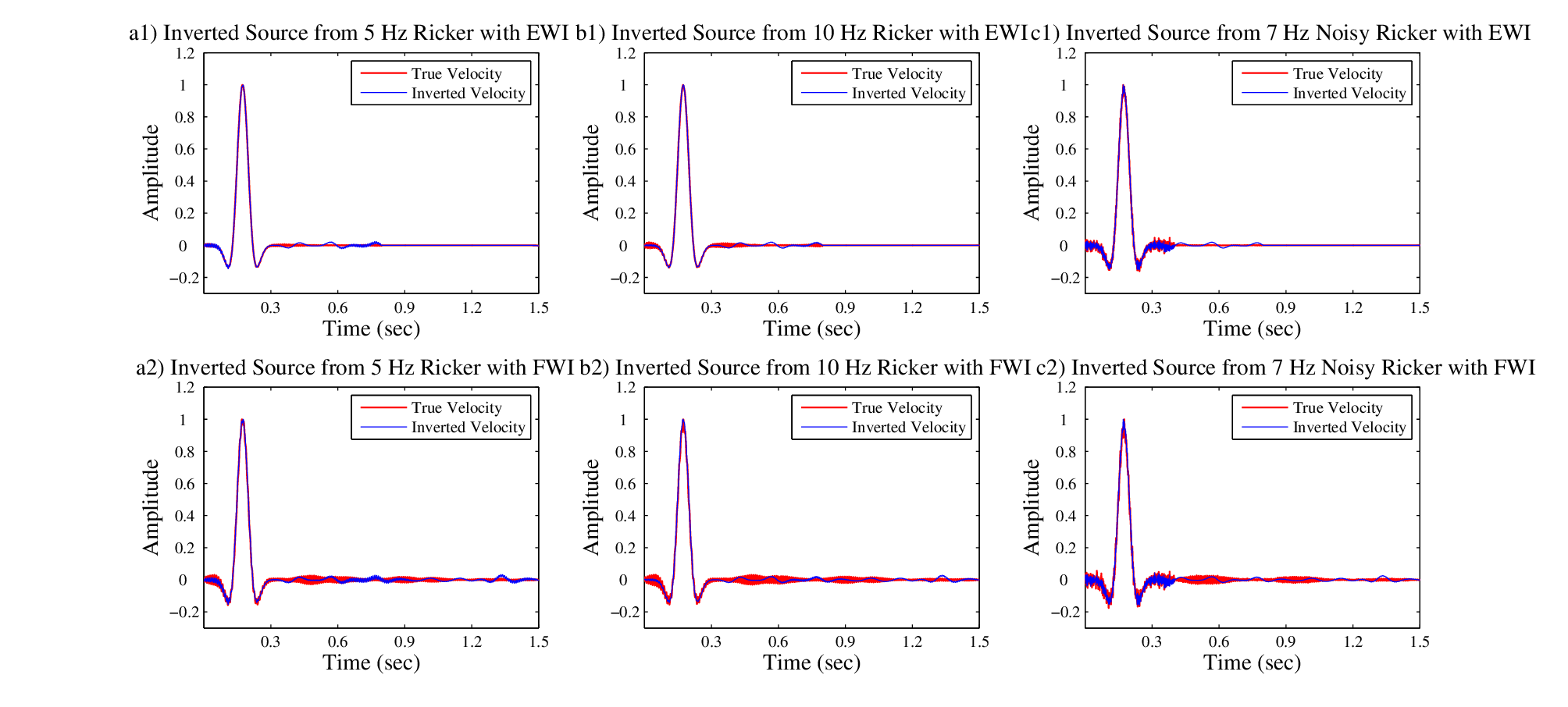}
    \caption{Sources inverted with true velocity models from three sources using EWI (a1, b1, and c1) and FWI (a2, b2, and c2), respectively. }
\label{fig12_1DInvSrcs_EWIFWI}
\end{figure*}

\begin{figure*}
\centering
\includegraphics[width=4.5in]{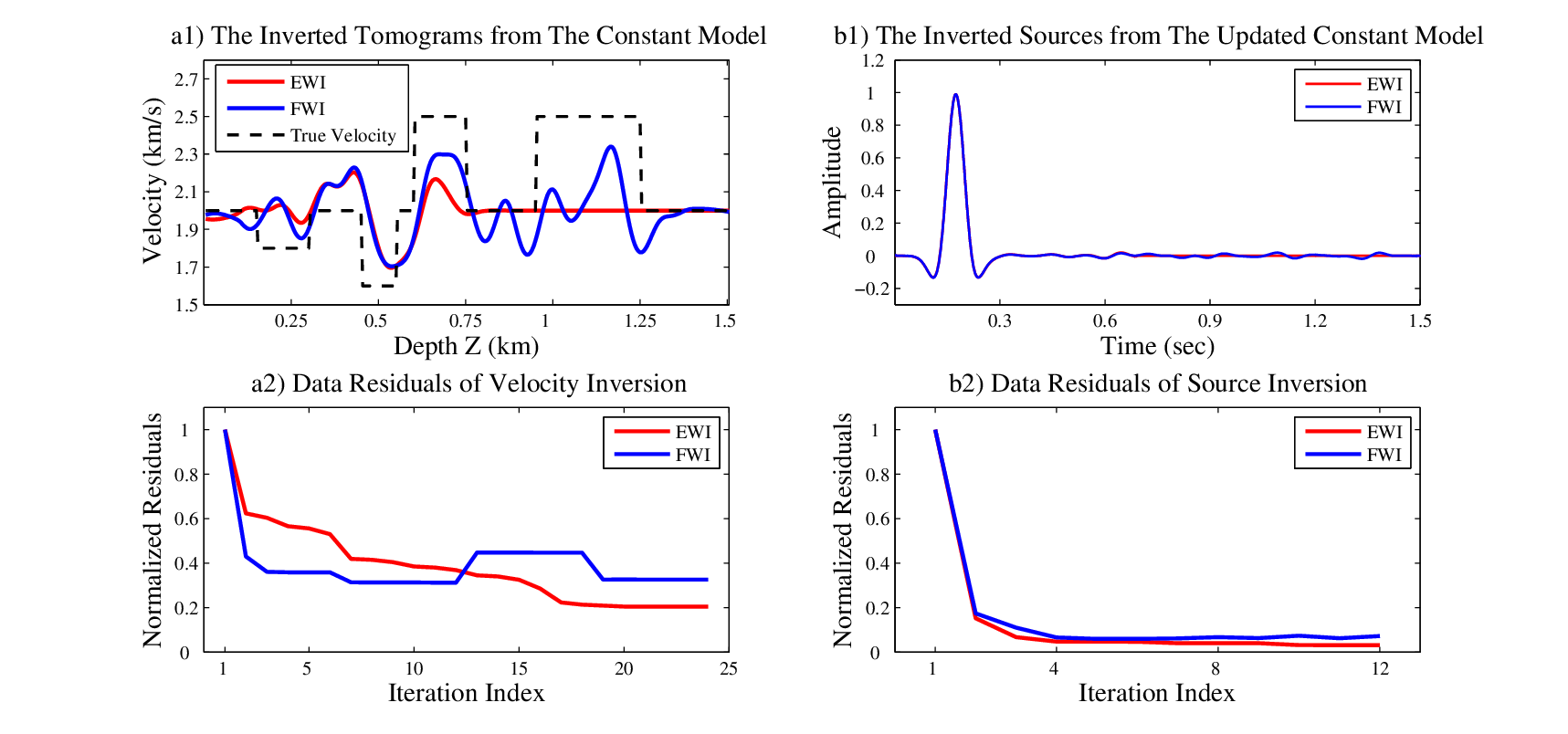}
    \caption{ Collocated inversions of (a) waveforms and (b) sources alternatively with Ricker source peaked at 5 Hz from the constant velocity model using EWI and FWI, and their associated data residuals (c) and (d).  }
\label{fig13_1DsrcEWI_constModel_5hz}
\end{figure*}

\begin{figure*}
\centering
\includegraphics[width=4.5in]{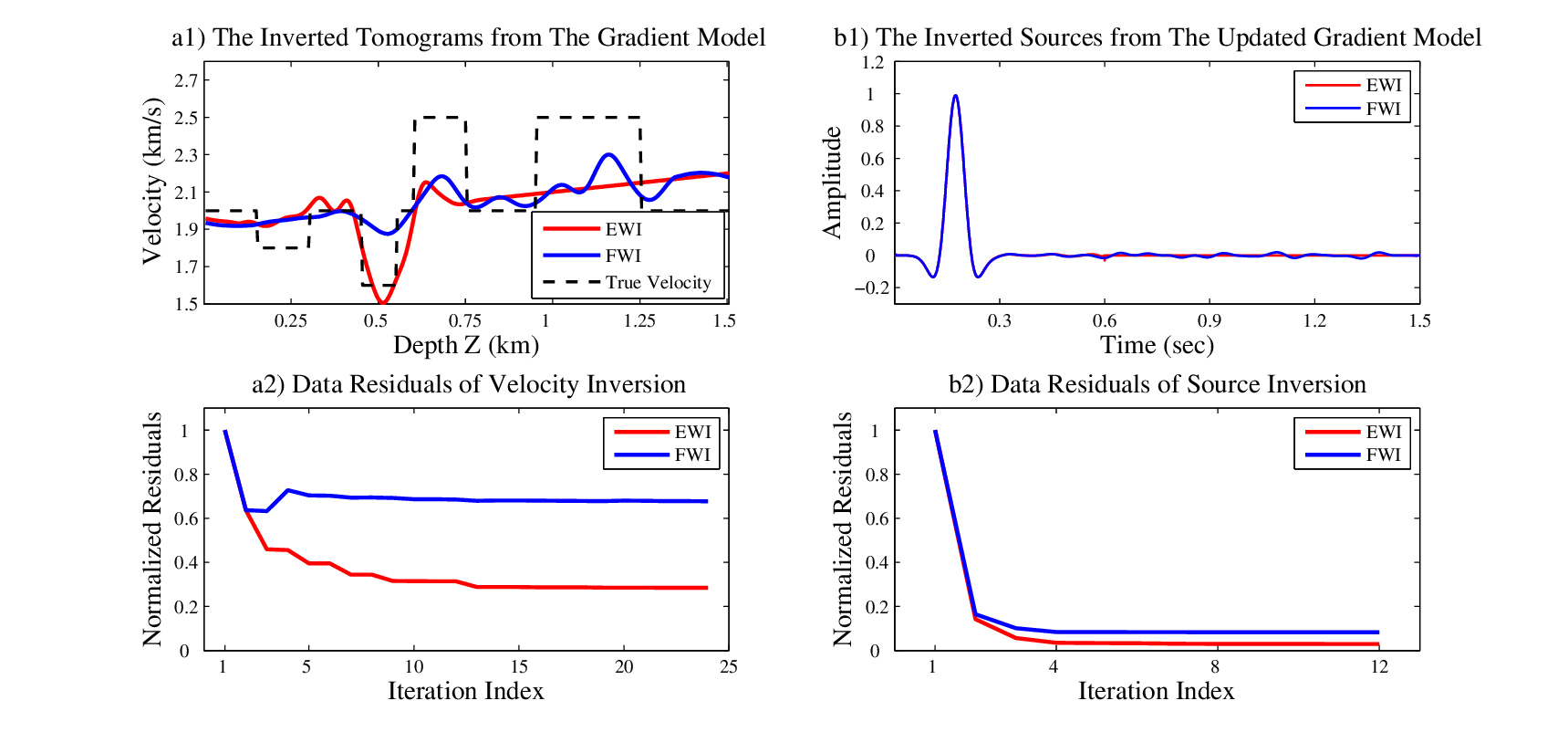}
    \caption{Collocated inversions of (a) waveforms and (b) sources alternatively with the Ricker source peaked at 5 Hz from the gradient velocity model using EWI and FWI, and their associated data residuals (c) and (d).   }
\label{fig14_1DsrcEWI_gradModel_5hz}
\end{figure*}

\begin{figure*}
\centering
\includegraphics[width=4.5in]{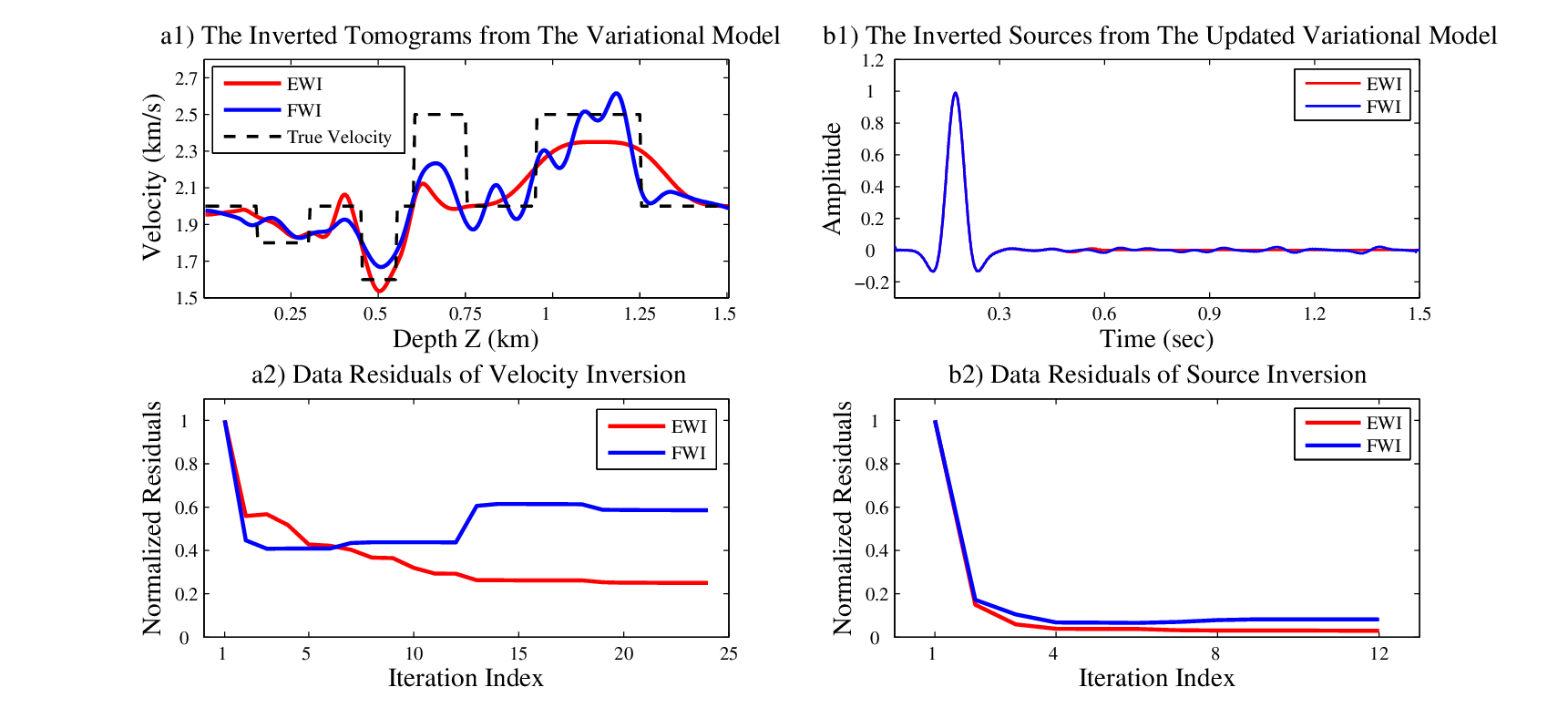}
    \caption{Collocated inversions of (a) waveforms and (b) sources alternatively with Ricker source peaked at 5 Hz from the variational velocity model using EWI and FWI, and their associated data residuals (c) and (d). }
\label{fig15_1DsrcEWI_varModel_5hz}
\end{figure*}

\begin{figure*}
\centering
\includegraphics[width=4.5in]{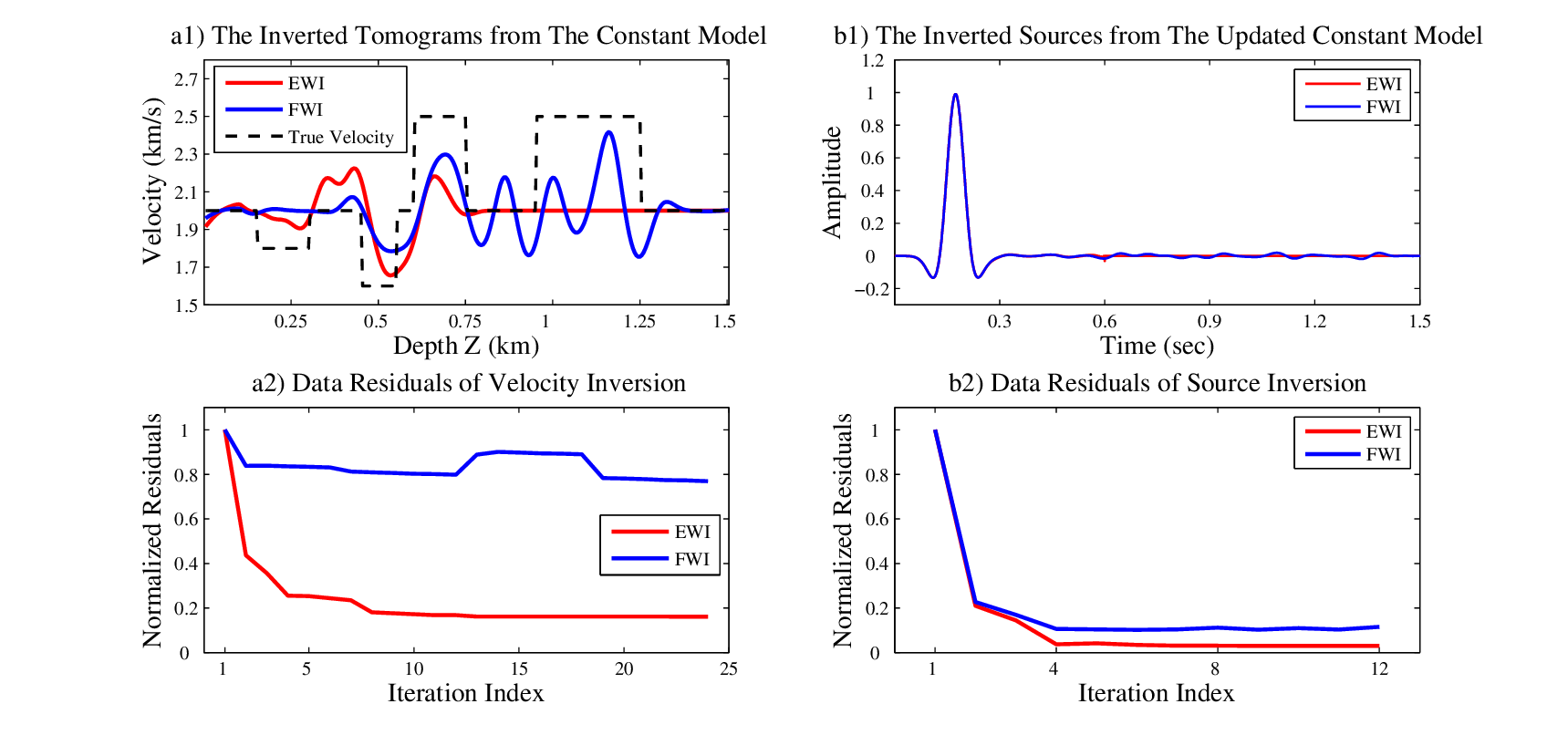}
    \caption{Collocated inversions of (a) waveforms and (b) sources alternatively with Ricker source peaked at 10 Hz from the constant velocity model using EWI and FWI, and their associated data residuals (c) and (d).   }
\label{fig16_1DsrcEWI_constModel_10hz}
\end{figure*}

\begin{figure*}
\centering
\includegraphics[width=4.5in]{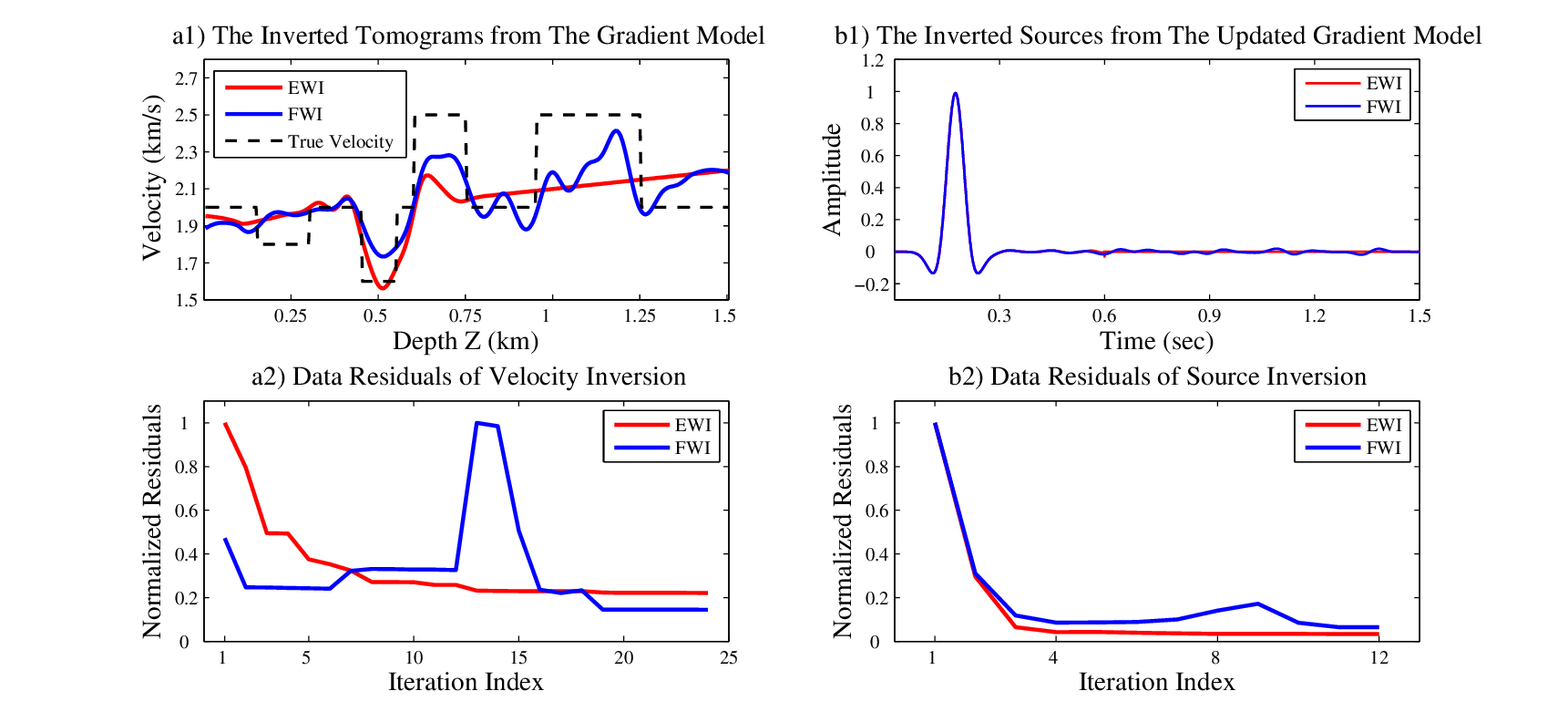}
    \caption{Collocated inversions of (a) waveforms and (b) sources alternatively with Ricker source peaked at 10 Hz from the gradient velocity model using EWI and FWI, and their associated data residuals (c) and (d).   }
\label{fig17_1DsrcEWI_gradModel_10hz}
\end{figure*}

\begin{figure*}
\centering
\includegraphics[width=4.5in]{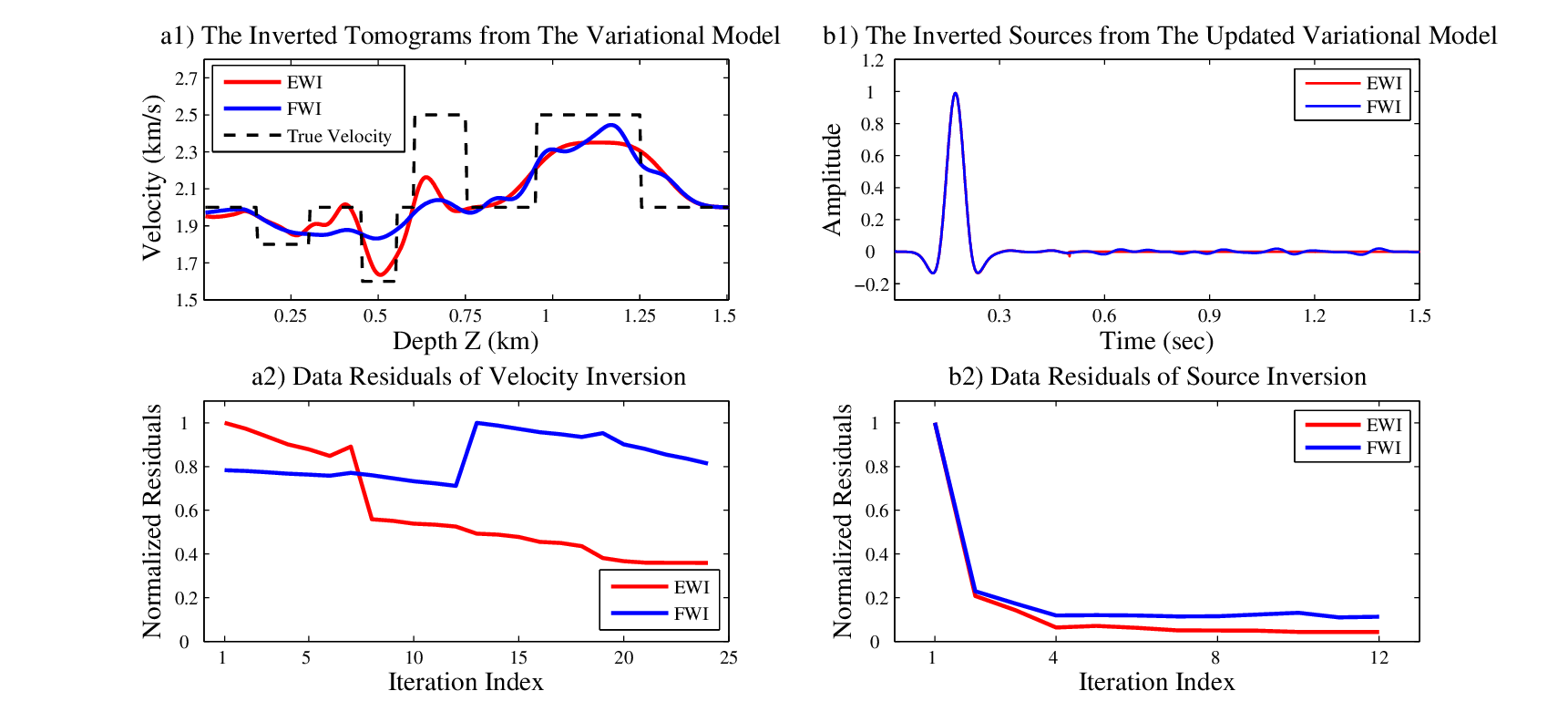}
    \caption{Collocated inversions of (a) waveforms and (b) sources alternatively with Ricker source peaked at 10 Hz from the variational velocity model using EWI and FWI, and their associated data residuals (c) and (d). }
\label{fig18_1DsrcEWI_varModel_10hz}
\end{figure*}

\begin{figure*}
\centering
\includegraphics[width=4.5in]{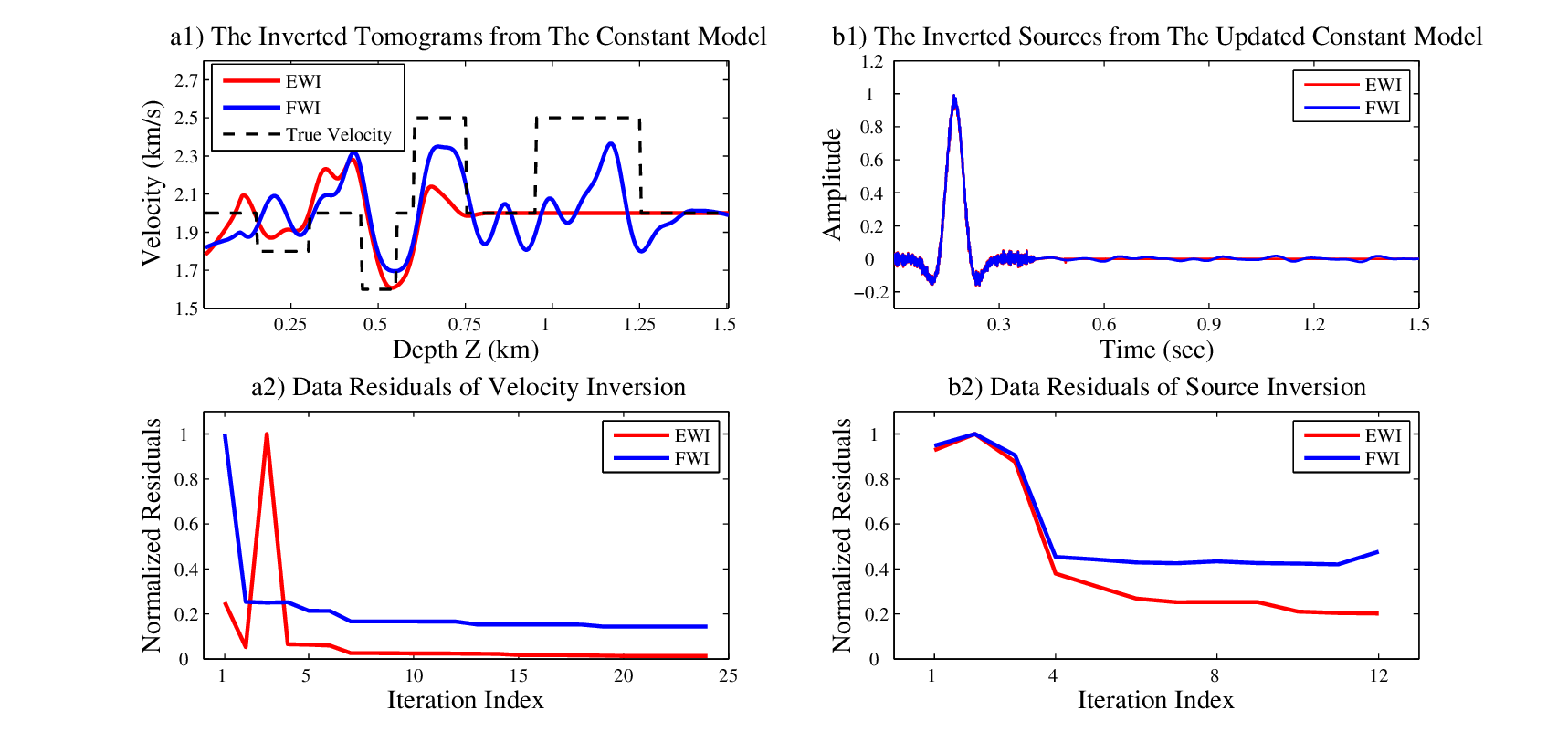}
    \caption{Collocated inversions of (a) waveforms and (b) sources alternatively with the strong noisy Ricker source in Fig. \ref{fig10_1DNoiseSrcs}(b) from the constant velocity model using EWI and FWI, and their associated data residuals (c) and (d). }
\label{fig19_1DsrcEWI_constModel_7hzm}
\end{figure*}

\begin{figure*}
\centering
\includegraphics[width=4.5in]{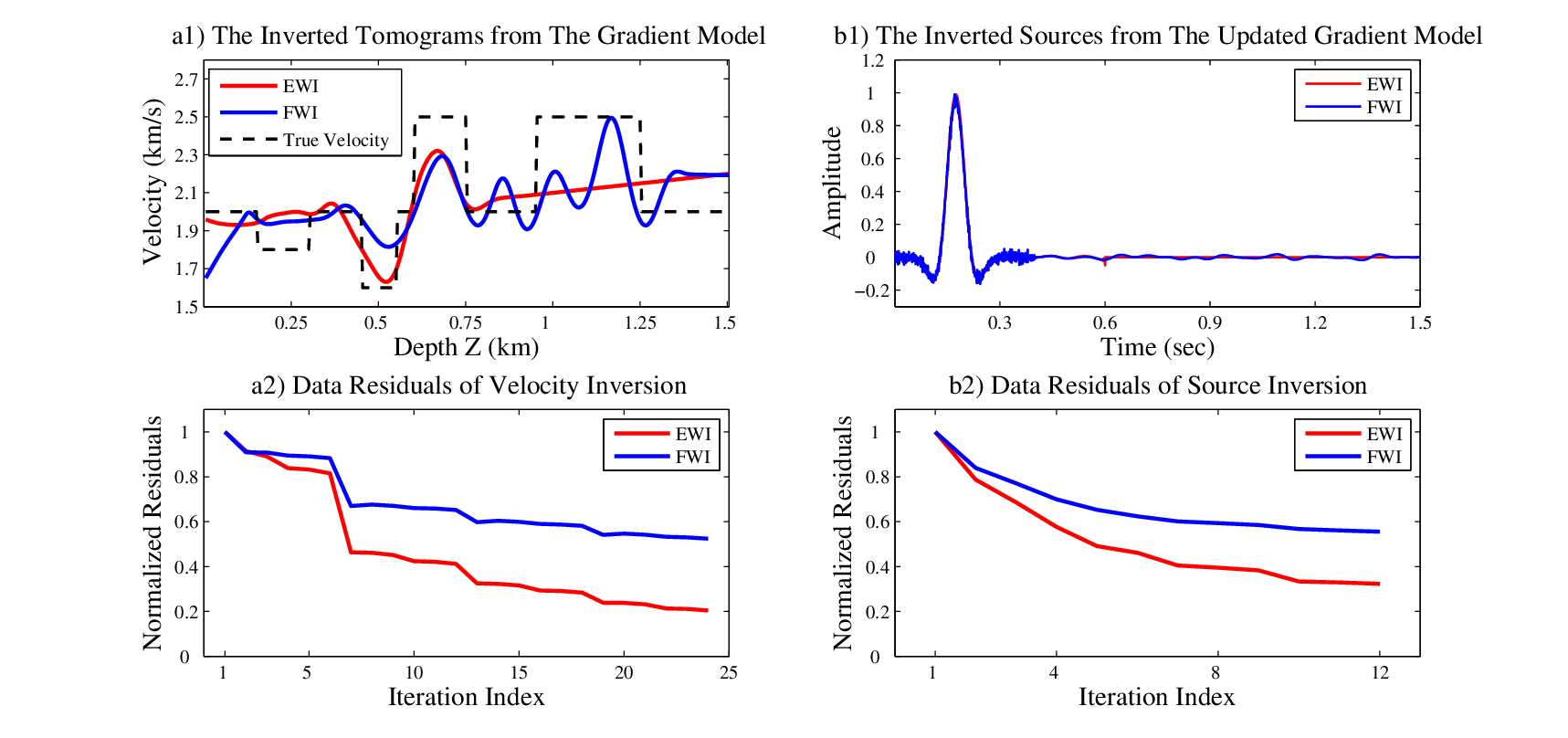}
    \caption{Collocated inversions of (a) waveforms and (b) sources alternatively with the strong noisy Ricker source in Fig. \ref{fig10_1DNoiseSrcs}(b) from the gradient velocity model using EWI and FWI, and their associated data residuals (c) and (d).}
\label{fig20_1DsrcEWI_gradModel_7hzm}
\end{figure*}

\begin{figure*}
\centering
\includegraphics[width=4.5in]{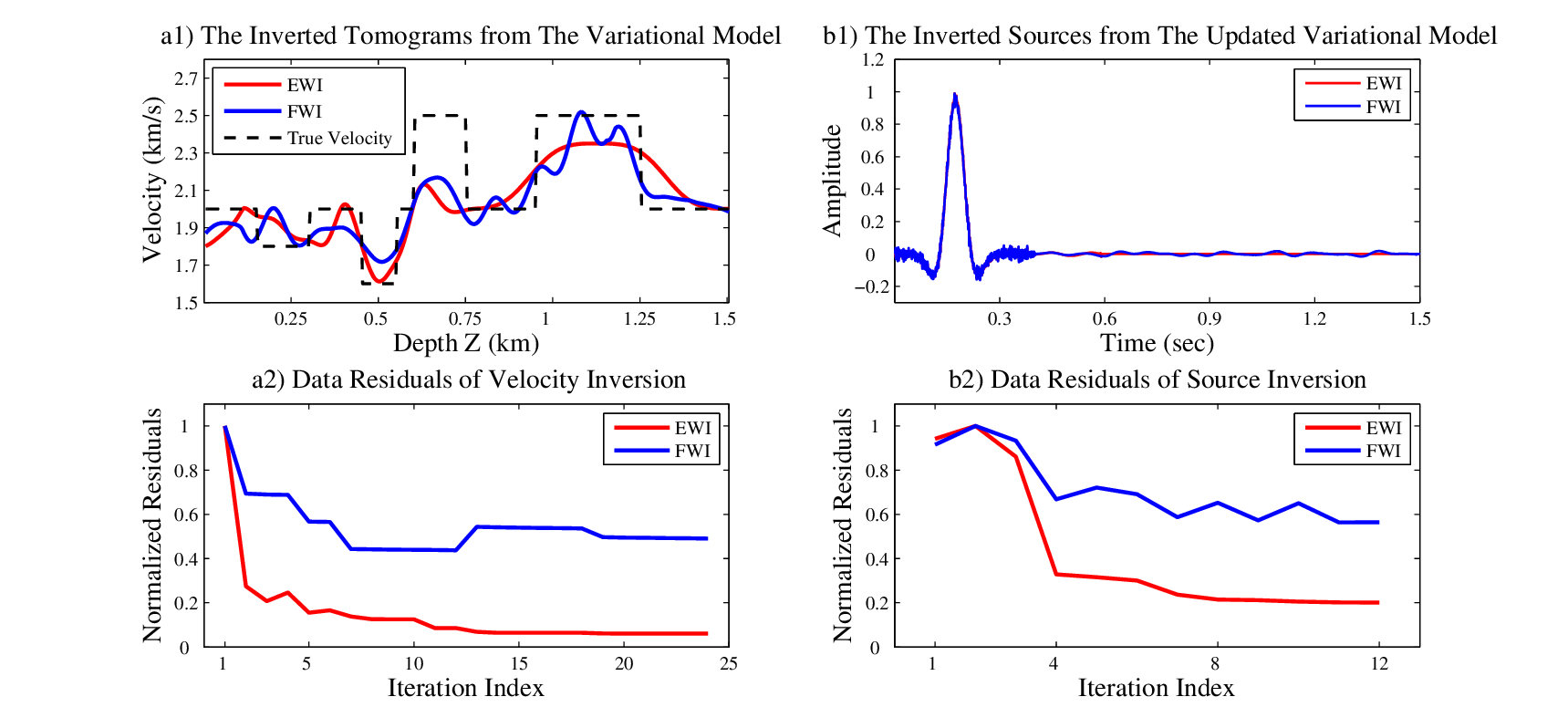}
    \caption{Collocated inversions of (a) waveforms and (b) sources alternatively with the strong noisy Ricker source in Fig. \ref{fig10_1DNoiseSrcs}(b) from the variational velocity model using EWI and FWI, and their associated data residuals (c) and (d).}
\label{fig21_1DsrcEWI_varModel_7hzm}
\end{figure*}

\indent The reasons to conduct one-dimensional examples can be summed up as follows. Firstly, the one-dimensional model, although uncommon in real practice, is not that straightforward to work on with only one source-receiver pair on the surface in our study.  Secondly, owing to the simple acquisition settings, such one-dimensional case becomes more pure and more helpful to unveil the properties of the collocated inversion without image stacking.
\\\indent Fig. \ref{fig2_1D_srcsAlign} shows the sources that are used to generate synthetic data and invert the two models in Fig. \ref{fig3_1DTrueModels}, whose sizes are all $1.5~km$ in the vertical direction with a grid point of spacing of 5 $m$. Fig. \ref{fig3_1DTrueModels}a is different from Fig. \ref{fig3_1DTrueModels}b only in the shallow part above $0.6~km$ in depth. One is with a layered near surface velocity distribution, and the other is with a gradient one. The data are recorded by just one receiver at the surface, and they are triggered by the source at the receiver's  location without the free surface condition, but with the absorbing boundary condition at the end. Based on the wave equation (\ref{eqn_acou}), a Ricker wavelet with a frequency band peaked at 7 $Hz$, namely the true source in Fig. \ref{fig2_1D_srcsAlign}, is excited to simulate the forward modeling, and two recorded seismograms are shown in Figs. \ref{fig4_1DSeismograms}a and \ref{fig4_1DSeismograms}b, corresponding to the two models in Fig. \ref{fig3_1DTrueModels}. The recording time for each of them is 1.5 $sec$ with a sampling interval of 1 $ms$. Therefore the number of temporal samples is set as 1500 for each trace. Three initial models, shown in Fig. \ref{fig5_1DInitialVel}, are used to test inversions. One is a constant velocity model, while another is a linearly increasing velocity model; the rest one is a smoothly variational model. According to the parameters set above, only reflections are recorded in either the early arrivals or the full waveforms. However, it is still effective to carry out the numerical experiments with the proposed theory in last section. Next, we first show the inversion results using the true source and incorrect sources, respectively. We then test how much the true source can be reconstructed from wrong sources with the true velocity model. Finally, results of the collocated inversion will be presented under mixed circumstances.
\\\indent To assess the proposed methods, two conventional methods, EWI and FWI, are respectively conducted on the three starting models (Fig. \ref{fig5_1DInitialVel}) with the true source. Then three control groups of results, treated as benchmarks, are shown in Figs. \ref{fig6_1DInvConstModels_trueSrc}-\ref{fig8_1DInvVarModels_trueSrc}. Here, we choose the time window from 0 to 1 $s$ for EWI so that only a few reflections can govern the inversion. Under these restrictions, only the velocity profiles in red above 0.8 $km$ can be inverted by EWI, compared to the blue profiles reconstructed by FWI. In most cases, it is apparent that EWI converges faster than FWI from the data residuals shown in Figs. (\ref{fig6_1DInvConstModels_trueSrc}-\ref{fig8_1DInvVarModels_trueSrc})(c-d). Moreover, near-surface velocity models can be better restored by EWI than by FWI.
\\\indent To carry out routine experiments using incorrect sources, the Ricker wavelets with frequency bands peaked at 5 $Hz$ and 10 $Hz$, are used to conduct EWI still with the three initial models. Unfortunately, neither of them (Figs. \ref{fig9_1DInv3Models_wrongSrcs}(a-b)) can provide even reluctantly satisfactory results after 20 iterations. To alleviate the influence of sources on EWI, two modified sources constructed by adding two types of noise (Figs. \ref{fig10_1DNoiseSrcs}a and \ref{fig10_1DNoiseSrcs}b) to the true source, are used for testing their performance.
The two noisy sources, although with the right peak frequency, can help EWI converge to some extent with their results shown in Fig. \ref{fig11_1DInv3Models_noiseSrcs}(a-b), whose associated normalized residuals are inferior to their noise free counterparts in Figs. \ref{fig6_1DInvConstModels_trueSrc}-\ref{fig8_1DInvVarModels_trueSrc}. It is also worthwhile to notice that the source with light noise, as expected, can obviously produce better results than that with strong noise. Therefore, we choose three incorrect sources, the two with peak frequencies at 5 $Hz$ and 10 $Hz$, and the strong noisy source, to investigate the source inversion procedure by both EWI and FWI according to the method in Section \ref{sec2A}. We fix two velocity models for further comparisons, the true one in Fig. \ref{fig3_1DTrueModels}a, and the EWI inverted one in Fig. \ref{fig8_1DInvVarModels_trueSrc}a.  Fig. \ref{fig12_1DInvSrcs_EWIFWI} presents the source inversion results by EWI in the first row, and by FWI in the second row. It is clear all the three input sources can be updated to approach the true one by either EWI or FWI in less than ten iterations, but obviously with superior results corresponding to EWI. Another phenomenon deserving more attention is that the true model may not help to produce better sources than the inverted model because the latter one should be smoother, hence raising fewer artifacts.
\\\indent Finally, the updated sources and velocity models cooperate with each other to complete the collocated inversion iteratively until convergence. Similar to the source inversion above, the three incorrect wavelets, together with the three starting models, are combined one-to-one as the initial parameters (see Fig. \ref{fig1_workFlow}) for the entire inversion loop, hence forming nine different input source-model pairs. Figs. \ref{fig13_1DsrcEWI_constModel_5hz}-\ref{fig15_1DsrcEWI_varModel_5hz} and Figs. \ref{fig16_1DsrcEWI_constModel_10hz}-\ref{fig18_1DsrcEWI_varModel_10hz} are two classes of inverted results with two initial sources, whose frequency bands are peaked at 5 $Hz$ and 10 $Hz$, respectively, while Figs. \ref{fig19_1DsrcEWI_constModel_7hzm}-\ref{fig21_1DsrcEWI_varModel_7hzm} are the inverted results with the noisy source in Fig. \ref{fig10_1DNoiseSrcs}b. We fix the rounds of iterations of the proposed collocated inversion by setting up 4 outer loops, and inside each loop, sources are updated first for 3 times and velocity models are updated later for 6 times by EWI or FWI. Therefore, as shown in Figs. \ref{fig13_1DsrcEWI_constModel_5hz}-\ref{fig21_1DsrcEWI_varModel_7hzm}(a2-b2), the total numbers of iterations are 12 and 24 for sources and velocity models, respectively. These nine figures indicate that the collocated inversion assisted by EWI shows more robust, accurate (for the near surface parts) and consistent results than the one by FWI, in terms of both the tomograms and the recovered sources. Therefore, the collocated inversion with early arrivals should be on the right track.
\subsection{Two-Dimensional Examples}
\begin{figure*}
\centering
\includegraphics[width=4.5in]{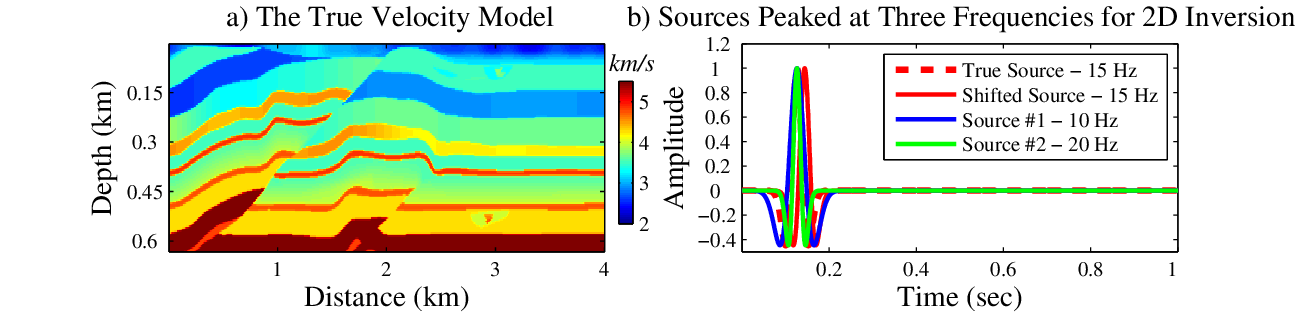}
    \caption{(a) True velocity model - a vertical slice of the SEG/EAGE overthrust model. (b) Four Ricker sources for data producing and inversion.}
\label{fig22_2DTrueModelSrcs}
\end{figure*}

\begin{figure*}
\centering
\includegraphics[width=4.5in]{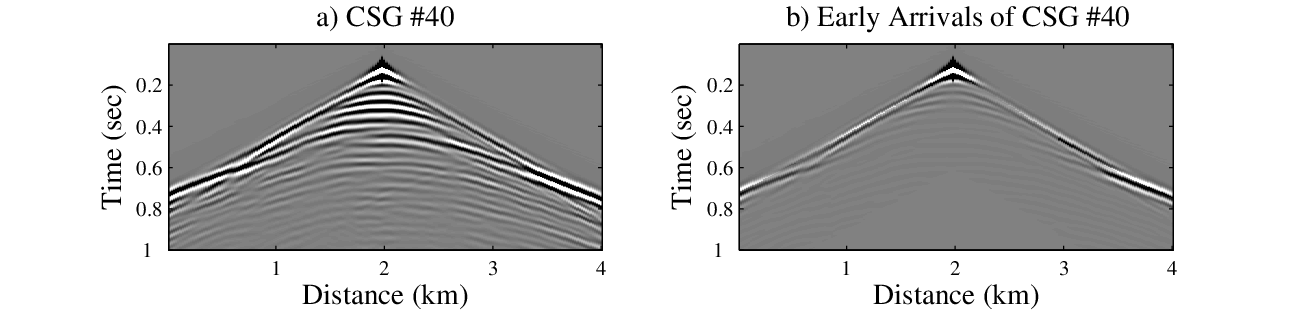}
    \caption{(a) The $40^{th}$ CSG corresponding to Fig. \ref{fig22_2DTrueModelSrcs}(a), and its (b) early arrivals. }
\label{fig23_2DTrueCSGEarr}
\end{figure*}

\begin{figure*}
\centering
\includegraphics[width=4.5in]{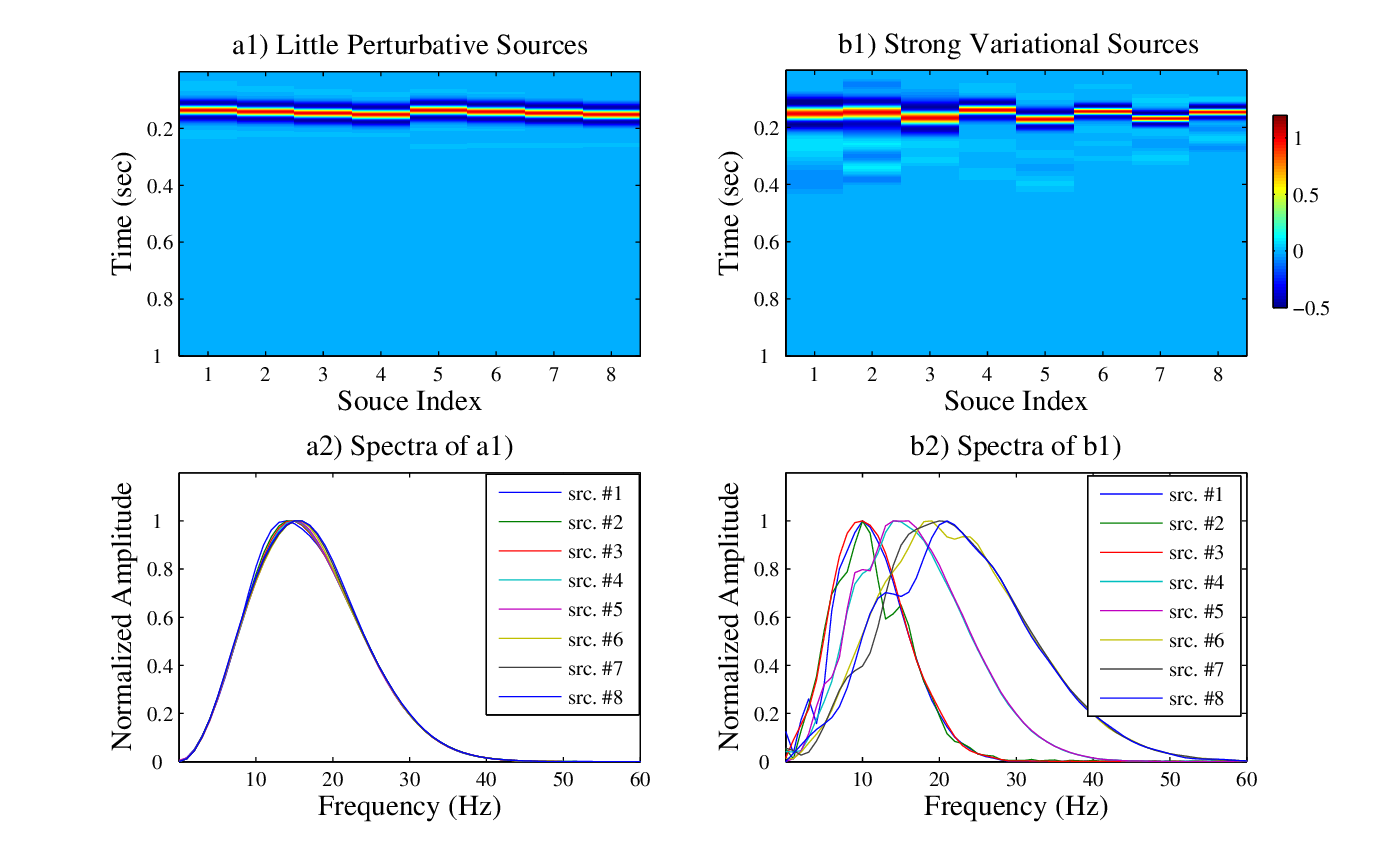}
    \caption{Two groups of sources to produce two data sets for inversion, one with (a1) little perturbative sources whose spectra are presented in (a2), and the other with (b1) strong variational sources whose spectra are presented in (b2). }
\label{fig24_2DvarstrSrcs}
\end{figure*}

\begin{figure*}
\centering
\includegraphics[width=4.5in]{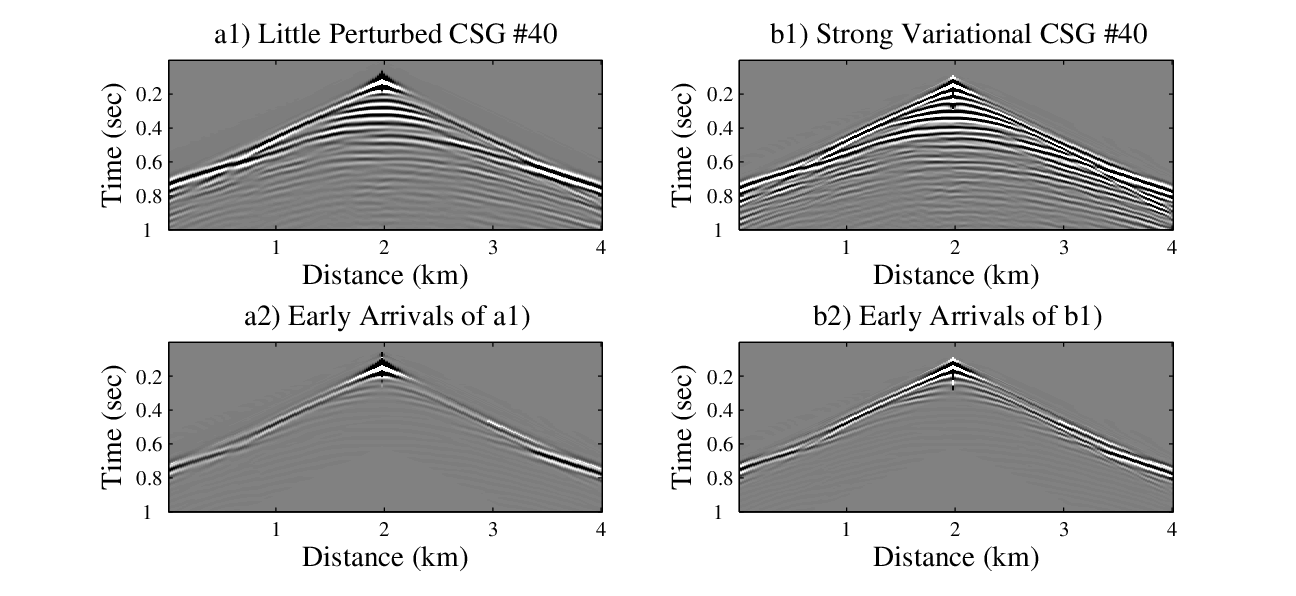}
    \caption{ (a1) CSG \#40 generated by Fig. \ref{fig24_2DvarstrSrcs}(a1) and its (a2) early arrivals, (b1) CSG \#40 generated by Fig. \ref{fig24_2DvarstrSrcs}(b1) and its (b2) early arrivals.}
\label{fig25_2Dseis_varstrEarr}
\end{figure*}

\begin{figure*}
\centering
\includegraphics[width=2.5in]{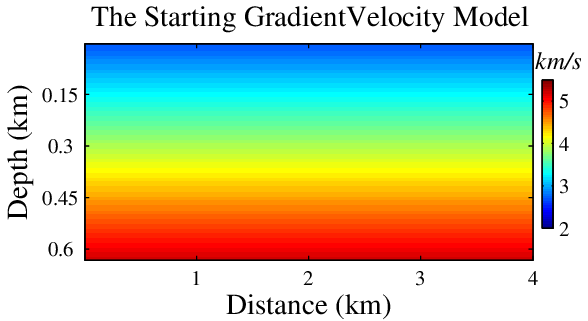}
    \caption{ The initial gradient velocity model.}
\label{fig26_2DInitModel}
\end{figure*}

\begin{figure*}
\centering
\includegraphics[width=4.5in]{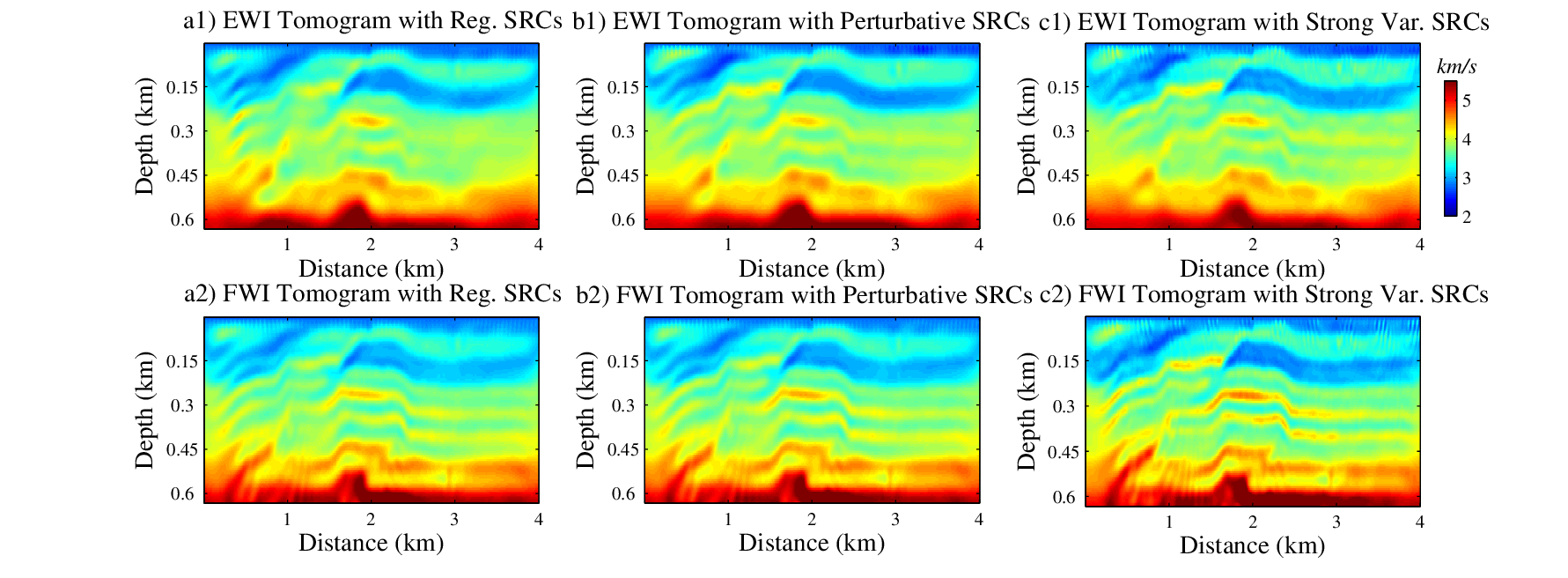}
    \caption{ Tomograms inverted respectively by EWI and FWI with (a1, a2) the true regular source, (b1, b2) the true perturbative sources in Fig. \ref{fig24_2DvarstrSrcs}(a1), and (c1, c2) the strong variational sources in Fig. \ref{fig24_2DvarstrSrcs}(b1). }
\label{fig27_2DInvModel_trueSrcs}
\end{figure*}

\begin{figure*}
\centering
\includegraphics[width=4.5in]{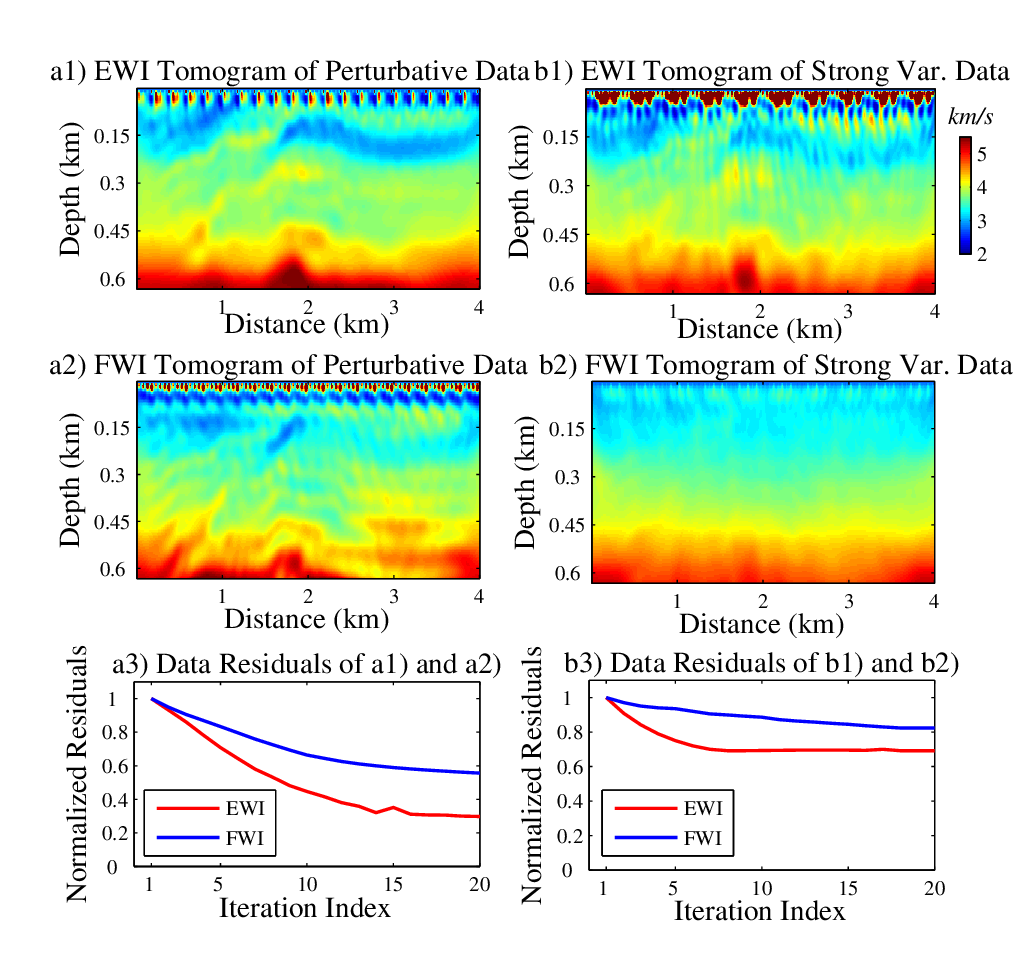}
    \caption{Tomograms respectively inverted by EWI and FWI with the shifted source peaked at 15 Hz in Fig. \ref{fig22_2DTrueModelSrcs}(b), (a1, a2) on the data set produced by the perturbative sources, and (b1, b2) on the data set produced by the strong variational sources. Their associated data residuals are presented in (a3) and (b3). }
\label{fig28_2DfailInvModels}
\end{figure*}

\begin{figure*}
\centering
\includegraphics[width=4.5in]{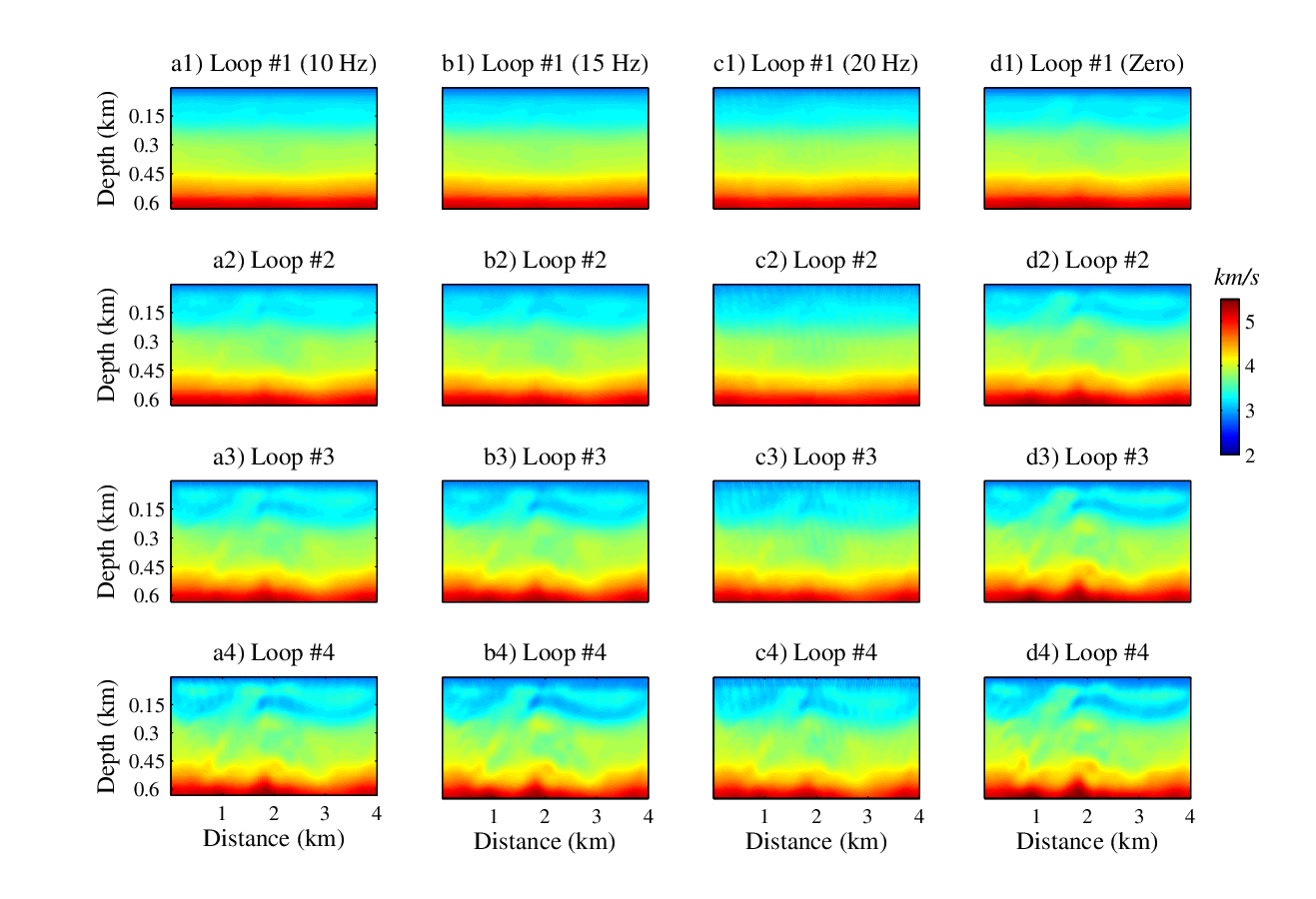}
    \caption{For the data set produced by the perturbative sources, updated tomograms for the first four iterations of the collocated inversion with the sources peaked at (a1-a4) 10 Hz, (b1-b4) 15 Hz, (c1-c4) 20 Hz, and (d1-d4) zero.}
\label{fig29_2DewiTomo_varSrcIter}
\end{figure*}

\begin{figure*}
\centering
\includegraphics[width=4.5in]{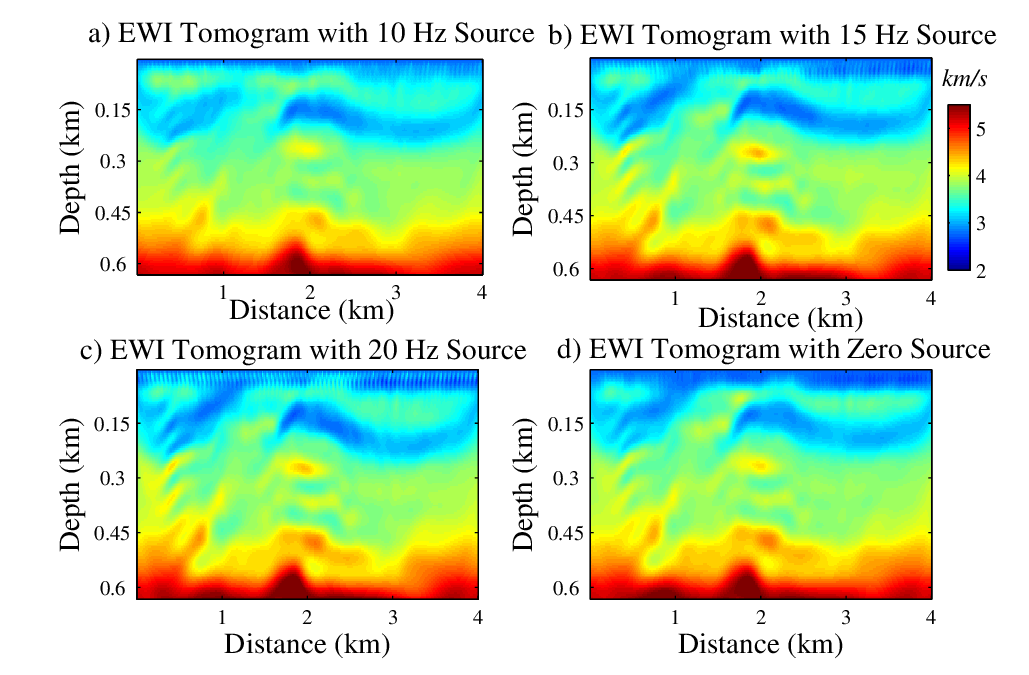}
    \caption{EWI tomograms of the collocated inversions with the sources peaked at (a) 10 Hz, (b) 15 Hz, (c) 20 Hz, and (d) zero.   }
\label{fig30_2DfinalTomo_srcInv_varSrc}
\end{figure*}

\begin{figure*}
\centering
\includegraphics[width=4.5in]{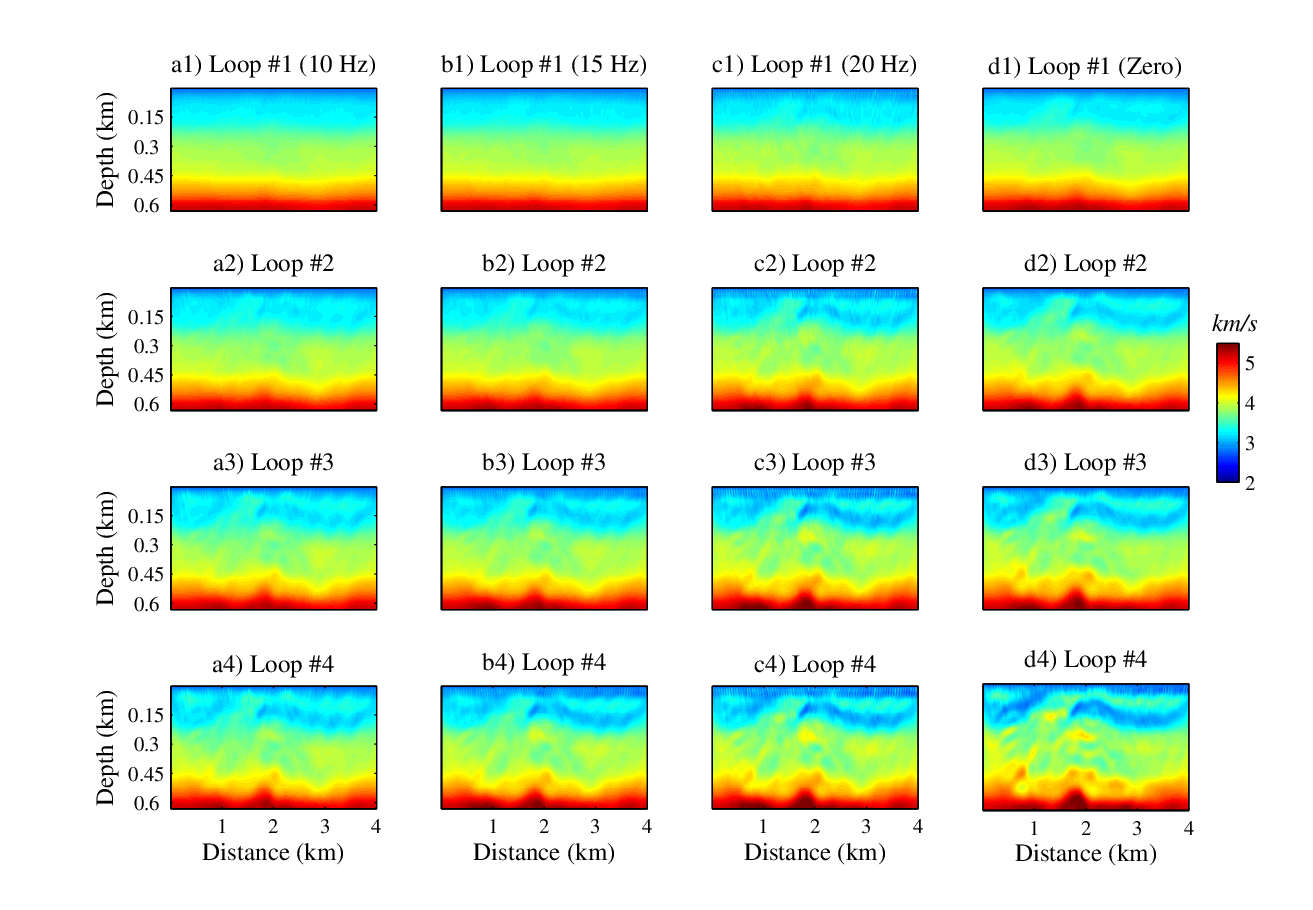}
    \caption{For the data set produced by the strong variational sources, updated tomograms for the first four iterations of the     collocated inversion with the sources peaked at (a1-a4) 10 Hz, (b1-b4) 15 Hz, (c1-c4) 20 Hz, and (d1-d4) zero.   }
\label{fig31_2DewiTomo_strSrcIter}
\end{figure*}

\begin{figure*}
\centering
\includegraphics[width=4.5in]{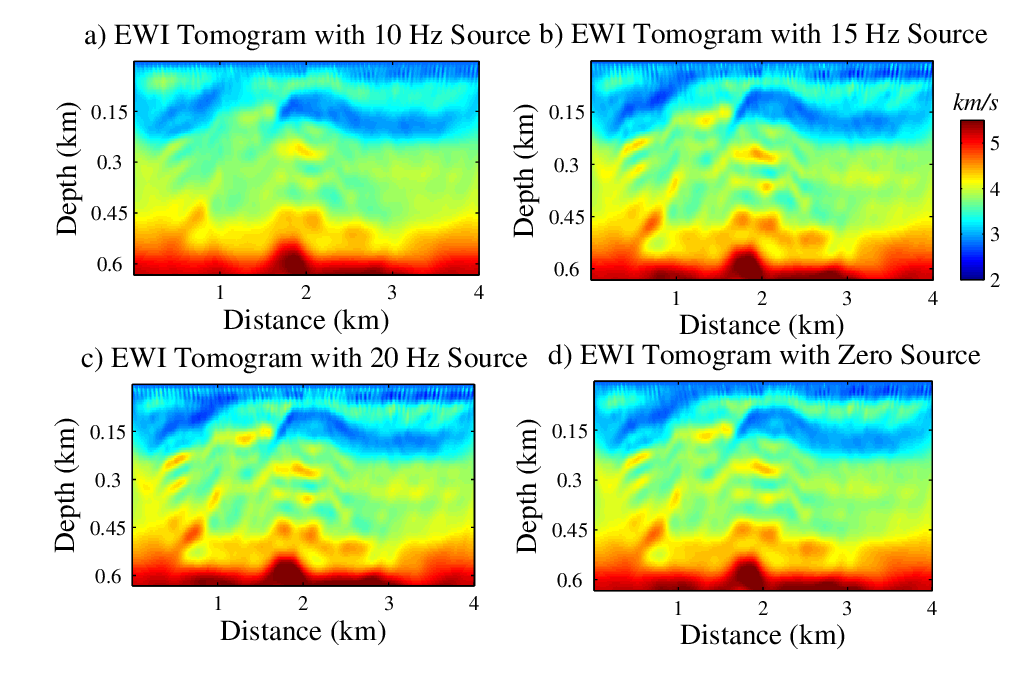}
    \caption{ EWI tomograms of the collocated inversions with the sources peaked at (a) 10 Hz, (b) 15 Hz, (c) 20 Hz, and (d) zero.  }
\label{fig32_2DfinalTomo_srcInv_strSrc}
\end{figure*}

\begin{figure*}
\centering
\includegraphics[width=4.5in]{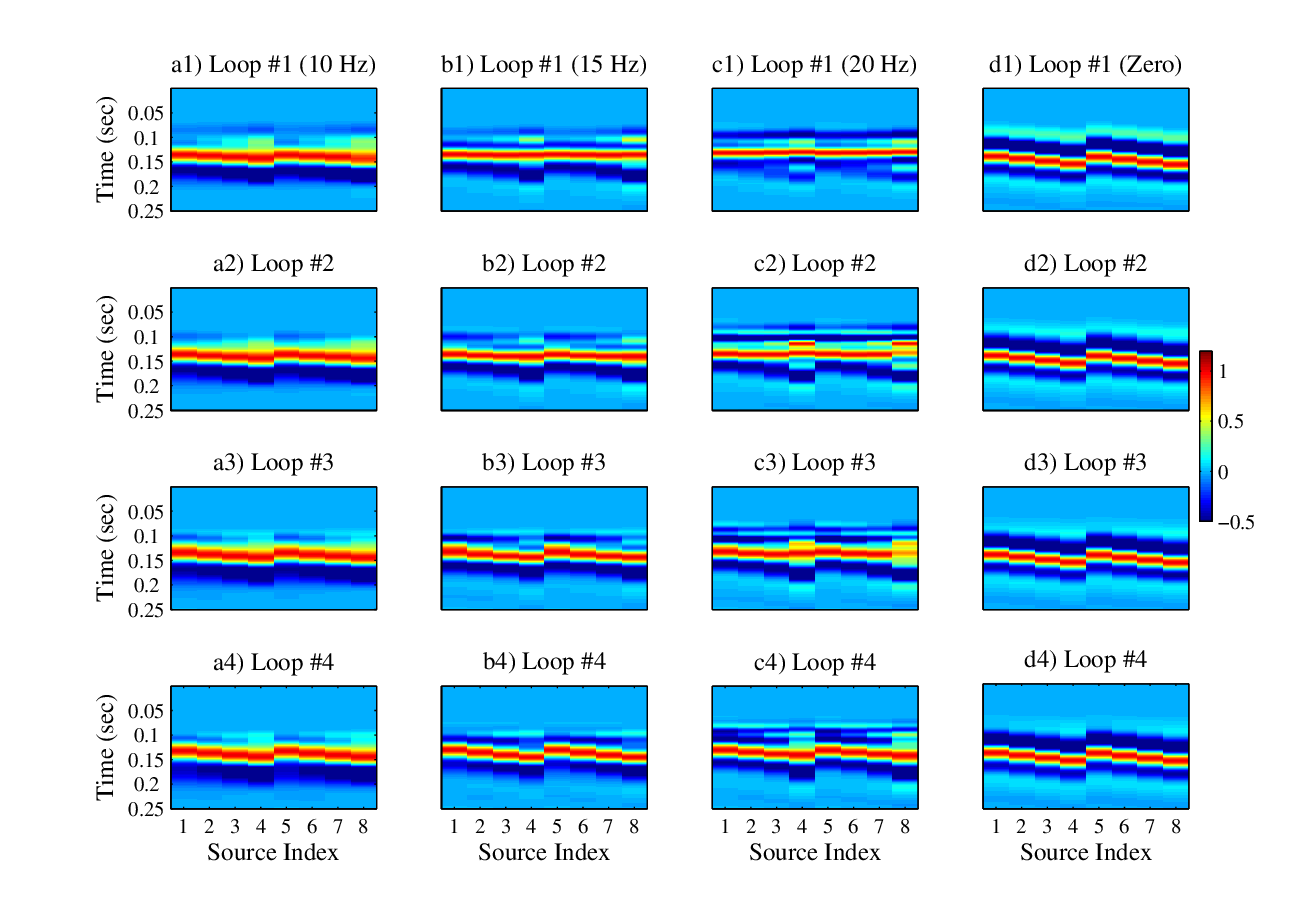}
    \caption{For the data set produced by the perturbative sources, updated sources for the first four iterations of the collocated inversion with the sources peaked at (a1-a4) 10 Hz, (b1-b4) 15 Hz, (c1-c4) 20 Hz, and (d1-d4) zero. }
\label{fig33_2DsrcInv_varSrcIter}
\end{figure*}

\begin{figure*}[!ht]
\centering
\includegraphics[width=4.5in]{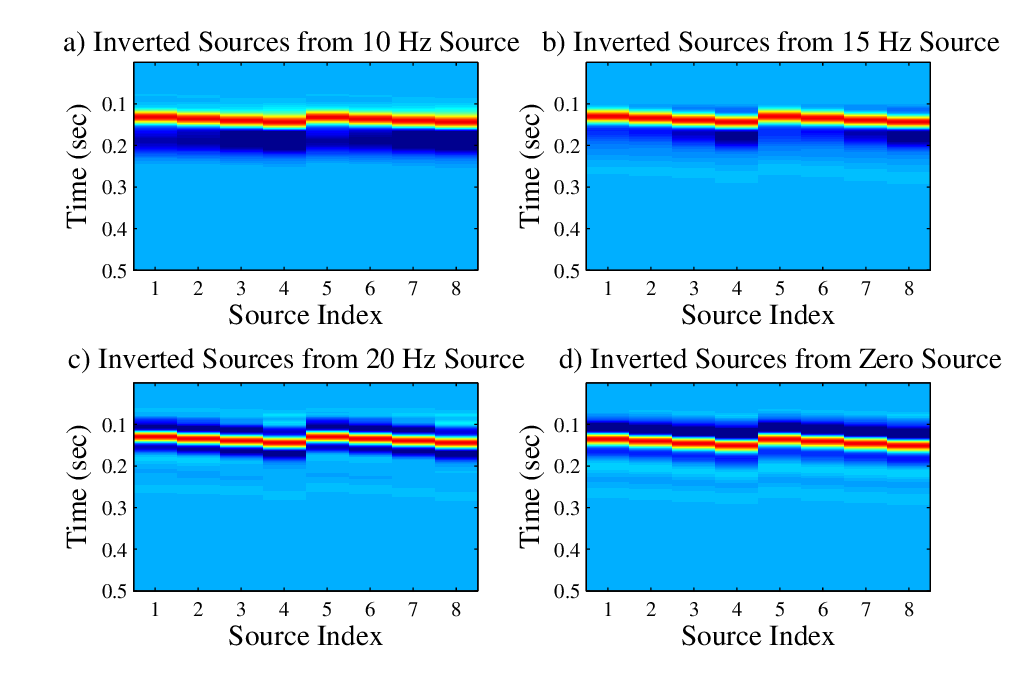}
    \caption{ Inverted sources of the collocated inversions with the sources peaked at (a) 10 Hz, (b) 15 Hz, (c) 20 Hz, and (d) zero.}
\label{fig34_2DfinalSrc_ewi_srcInv_varSrc}
\end{figure*}

\begin{figure*}[!ht]
\centering
\includegraphics[width=4.5in]{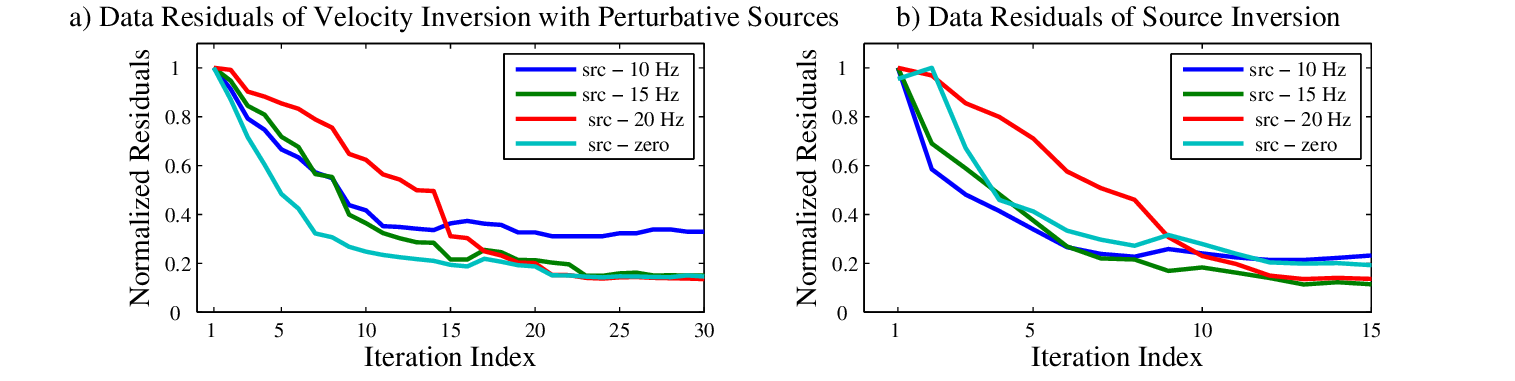}
    \caption{Data residuals of (a) EWI and (b) the source inversion embedded in the convergent collocated inversion for the data set produced by the perturbative sources. }
\label{fig35_2Dres_ewi_varSrc}
\end{figure*}

\begin{figure*}
\centering
\includegraphics[width=4.5in]{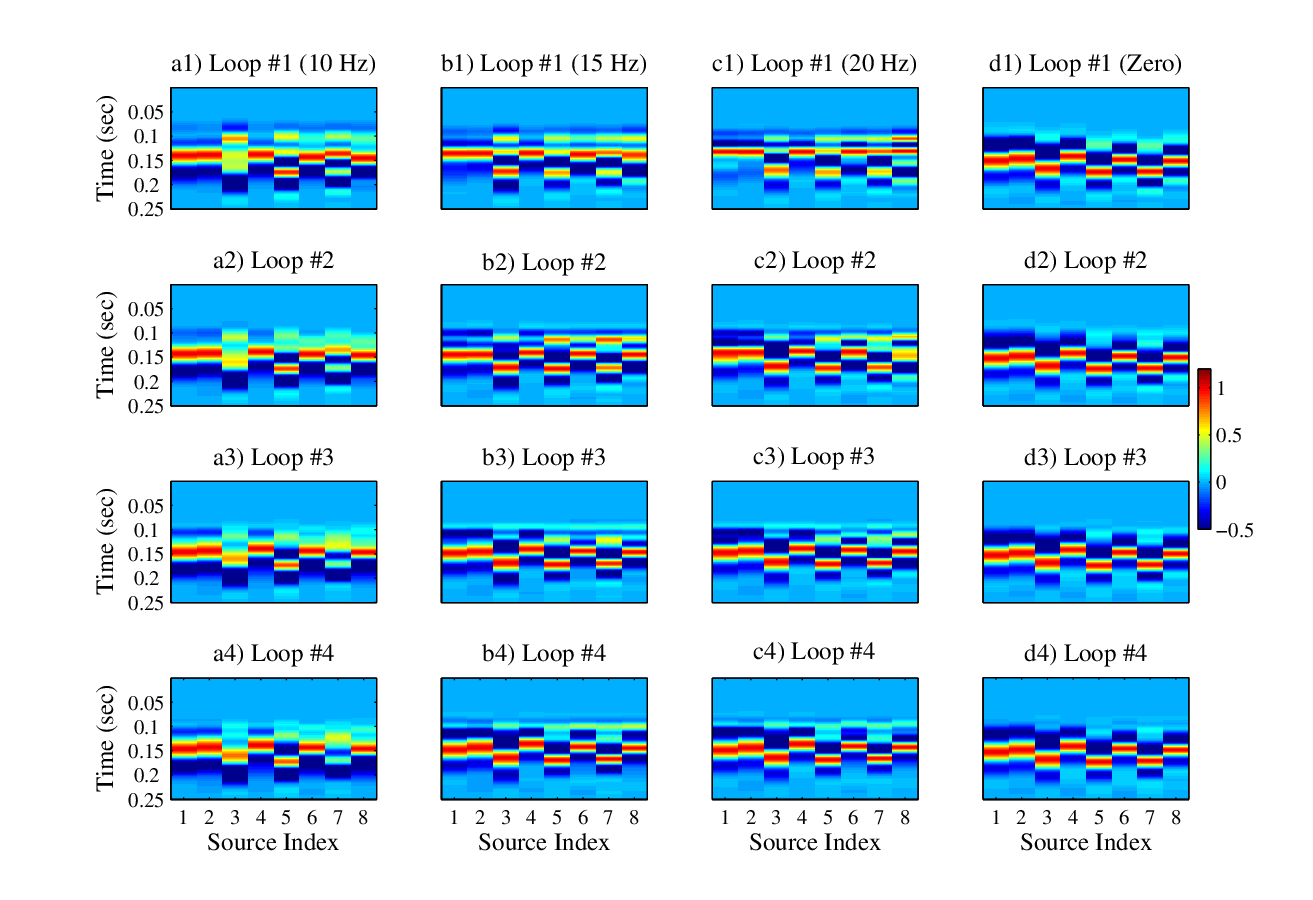}
    \caption{For the data set produced by the strong variational sources, updated sources for the first four iterations of the collocated inversion with the sources peaked at (a1-a4) 10 Hz, (b1-b4) 15 Hz, (c1-c4) 20 Hz, and (d1-d4) zero. }
\label{fig36_2DsrcInv_strSrcIter}
\end{figure*}

\begin{figure*}
\centering
\includegraphics[width=4.5in]{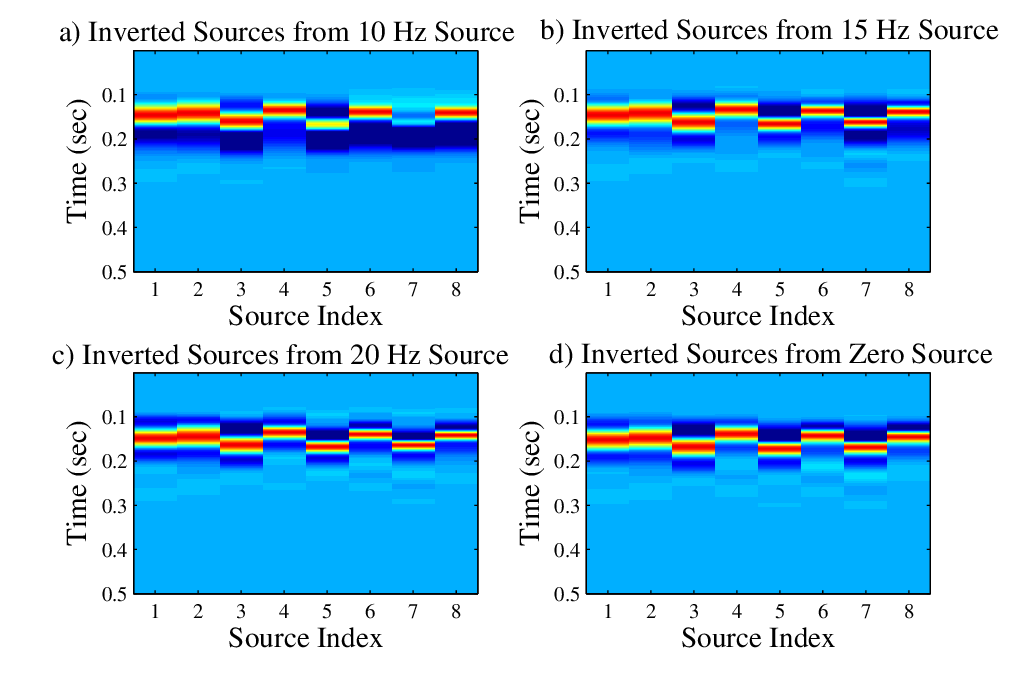}
    \caption{Inverted sources of the collocated inversions with the sources peaked at (a) 10 Hz, (b) 15 Hz, (c) 20 Hz, and (d) zero.}
\label{fig37_2DfinalSrc_ewi_srcInv_strSrc}
\end{figure*}

\begin{figure*}
\centering
\includegraphics[width=4.5in]{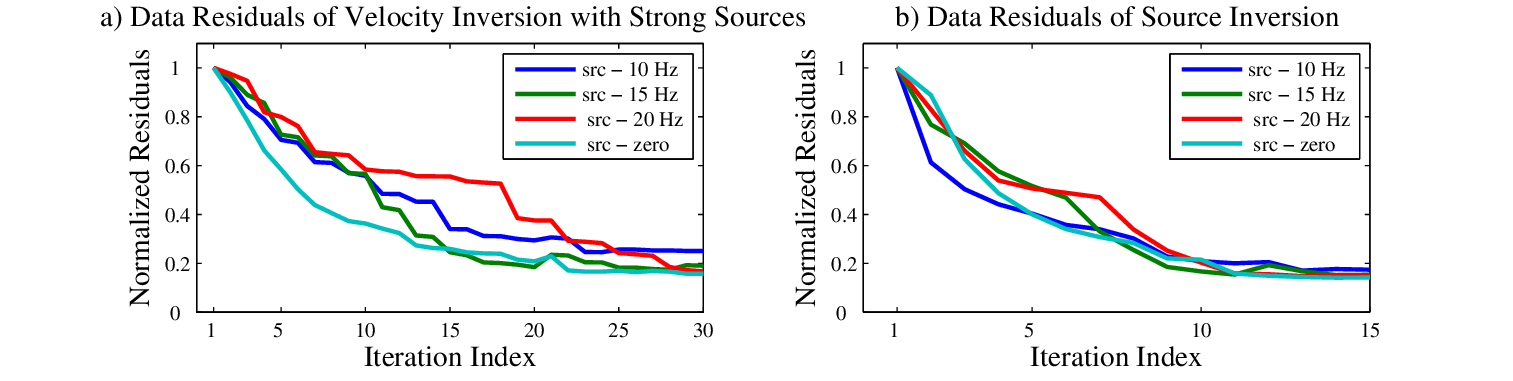}
    \caption{Data residuals of (a) EWI and (b) the source inversion embedded in the convergent collocated inversion for the data set produced by the strong variational sources. }
\label{fig38_2Dres_ewi_strSrc}
\end{figure*}

\indent To further verify the proposed method, two numerical experiments are conducted on a two-dimensional model for convincible results. A series of synthetic tests on a Society of Exploration Geophysicists (SEG)/European Association of Geoscientists and Engineers (EAGE) benchmark model will be presented, with two typical groups of multiple sources for imitating the real situation in real world.
\\\indent The model presented in Fig. \ref{fig22_2DTrueModelSrcs} is a vertical 2-D slice of the typical SEG/EAGE overthrust 3-D model, which is then used to generate several groups of synthetic data sets for validating the effectiveness of the proposed method. The model size is 4 $km~\times$ 0.63 $km$ in the horizontal and vertical directions, respectively. The signals are recorded by 200 receivers spaced at an interval of 20 $m$, and they are triggered by 80 sources spaced at an interval of 50 $m$. A grid point of spacing 5 $m$ is set for building the meshes with squared grids. Based on the acoustic wave equation (\ref{eqn_acou}), a Ricker wavelet with its frequency band peaked at 15 $Hz$, represented by the dashed red curve in Fig. \ref{fig22_2DTrueModelSrcs}b, excites as the reference source to simulate the forward modeling for all the regular CSGs. A typical CSG \#40 with its early arrivals are shown in Fig. \ref{fig23_2DTrueCSGEarr}, whose recording length in time is 1 $s$ with a sampling interval of 0.5 $ms$.
\\\indent However, in the field work, it is unrealistic to obtain CSGs with only one source, even if controllable vibrators or air guns are adopted. For this purpose, two groups of sources, shown in Fig. \ref{fig24_2DvarstrSrcs}(a1-b1) with 8 wavelets in each group, are constructed to test the collocated inversion method, and their spectra are presented in Fig. \ref{fig24_2DvarstrSrcs}(a2-b2). Each wavelet serves as the source for 10 CSGs, and they are evenly distributed along the acquisition line. So the same source reappears for every 400 $m$ in distance, or every 8 CSGs. We then use the two groups of sources to generate two new data sets, and it can be easily discerned that Fig. \ref{fig25_2Dseis_varstrEarr}(a1) is very different from Fig. \ref{fig25_2Dseis_varstrEarr}(b1). In particular, discrepancies of their early arrivals presented in Figs. \ref{fig25_2Dseis_varstrEarr}(a2) and \ref{fig25_2Dseis_varstrEarr}(b2) are also apparent. Here, we drop the 10 near-offset traces for inversion to avoid certain ``inversion crimes''. Intuitively, the collocated inversion could be more rewarding on the data set by the little perturbative sources because their frequency bands are almost all peaked at 15 $Hz$ just as the reference source, but with merely a few shifts in time or perturbations in amplitudes. However, when it comes to the data set by the strong variational sources, the entire inversion could be more complicated because their frequency bands are more complex. In other words, if only one fixed source is used to invert these two datasets, failures can be predicted in a great chance.
\\\indent To enhance comparisons, the three data sets generated above are respectively inverted with the initial gradient model in Fig. \ref{fig26_2DInitModel} and the true sources corresponding to them. As expected, all of them converge to reasonable results shown in Fig. \ref{fig27_2DInvModel_trueSrcs} by both EWI and FWI. However, even a small time shift added to the true reference source, shown by the red curve in Fig. \ref{fig22_2DTrueModelSrcs}b, can largely undermine the final results as indicated by Fig. \ref{fig28_2DfailInvModels}. Normalized residuals in Figs. \ref{fig28_2DfailInvModels}(a3-b3) imply that the perturbative data set should be easier to deal with than the the strong variational data set by merely the waveform inversion. Similar to the one-dimensional case, results of EWI are still more competitive than those of FWI for this 2-D model.
\\\indent Finally, according to the workflow in Fig. \ref{fig1_workFlow}, we execute the collocated inversion by setting up 15 outer loops, where sources and velocity models are respectively updated once and twice within every loop. Therefore, all the input sources are updated for 15 times while velocity models are updated for 30 times in total. Here, four initial sources, including the three wavelets (Fig. \ref{fig22_2DTrueModelSrcs}b) with peak frequencies at 10 $Hz$, 15 $Hz$, and 20 $Hz$, and the zero wavelet, take turns to participate in the collocate inversion. For the data set produced by the perturbative sources (Fig. \ref{fig24_2DvarstrSrcs}a), four intermediate models for the first eight iterations, actually corresponding to the four outer loops, are presented in Fig. \ref{fig29_2DewiTomo_varSrcIter}, and their final tomograms are shown in Fig. \ref{fig30_2DfinalTomo_srcInv_varSrc}. Similarly, for the data set produced by the strong variational sources (Fig. \ref{fig24_2DvarstrSrcs}b), the velocity models for the first four loops are presented in Fig. \ref{fig31_2DewiTomo_strSrcIter} with their final tomograms presented in Fig. \ref{fig32_2DfinalTomo_srcInv_strSrc}. Sometimes the so called natural sources are filtered out from the very near-offset traces, but these traces are discarded in our early settings so as not to make the inversion results too good to be true. This is the main reason we use zero source for comparisons; nevertheless, their back-propagated traces can still create the target wavelets that look superior to other initial sources, which may or may not coincide with the situation in real field cases. Comparisons of the inverted sources are also given as follows. The reconstructed little perturbative sources for the first four loops are presented in Fig. \ref{fig33_2DsrcInv_varSrcIter} with four columns corresponding to four different input sources, and their final results are shown in Fig. \ref{fig34_2DfinalSrc_ewi_srcInv_varSrc}. Data residuals of both EWI and the source inversion, shown in Figs. \ref{fig35_2Dres_ewi_varSrc}a and \ref{fig35_2Dres_ewi_varSrc}b, demonstrate positive feedbacks of their collocation. Accordingly, in the other experiment, the recovered strong variational sources for the first four loops are also presented in Fig. \ref{fig36_2DsrcInv_strSrcIter}, whose final results are shown in Fig. \ref{fig37_2DfinalSrc_ewi_srcInv_strSrc}. Their data residual curves are presented in Fig. \ref{fig38_2Dres_ewi_strSrc}. The collocated inversion, namely source inversion plus EWI, is more robust than only EWI in terms of producing credible background velocity models, because the sources are gradually updated to approximate equivalent ground truths.


\section{Conclusions and Discussions}
\indent We have presented a collocated inversion of sources and early arrival waveforms that can restore the true source wavelets and background velocity distributions alternatively. Numerical experiments in both the one-dimensional and two-dimensional cases demonstrate the effectiveness of the propose method compared with conventional EWI or FWI. The early arrival waveforms, usually with higher signal to noise ratio, are proved to be more reliable and suitable to this method. Therefore, EWI is recommended to be embedded in the collocated algorithms.
\\\indent An advantage of this approach is that it does not need an accurate near-surface velocity distribution, which can be adaptively approximated simply by repairing the input sources through fitting the early arrivals. Numerical experiments on the 1-D and the overthrust models validate its flexibilities of adapting to different initial models and various input sources, which can all be meliorated by complementing each other. Mutual benefits between the updated sources and velocity models can progressively guide the collocated inversion to convergence. In fact, one important reason to carry out quite a few numerical tests is to confirm the inverted tomograms under different circumstances. After all, an incorrect background velocity model is devastating no matter how well the sources can be inverted. However, from the perspective of source inversion, the back-propagated signals at the source location should be less polluted by EWI than by FWI due to the limitations of acquisition geometry. To stabilize the revised sources, taking the overthrust data set as an example, we update each source by averaging repaired sources from 10 CSGs because one source corresponds to ten gathers in our previous settings. In the field work, one CSG solely associated with one source should be recognized publicly in seismic exploration. Although this proposed extra process is easily conducted in calculation, it seems to add more computational costs than ordinary EWI. As a matter of fact, it is by our empirical experience that the sources should not be updated many times before the background velocity models become more reliable. A good ratio of source update and velocity update within one outer loop is either 1:2 or 1:3 to the best of our current knowledge.
\\\indent Looking back at the results above, liabilities of this method can be generalized in two aspects. One is that the properties of sources are largely determined by the data quality; and reversely, the other is that the calculated data set is very sensitive to small variations of the sources. To avoid such possible interferences, we cut off the very near-offset traces as well as the very far-offset traces in the overthrust case. That is another reason to explain why the zero wavelets behave quite well in the collocated inversion. Therefore, boosting the signal-to-noise ratio using seismic interferometry or muting unwanted noise with related techniques could be another necessary pre-processing step.
\\\indent Our future work needs to extend the application of this method to more real data for verifying the tomogram reliability and the stability of source inversion. No matter in traditional waveform inversion techniques, or in current machine learning aided methods, most research will assume that the sources are correct, which is not consistent with the real conditions in field data set. After all, if only one source is used for inverting a few CSGs, it will be a tough task full of trial and error. The possible successful strategy is to unify different technologies together and to graft them on EWI or FWI.

%

\begin{acknowledgments}
Chaiwoot Boonyasiriwat and Gerard Schuster have provided long-term selfless help on this material, and I would particularly like to thank them and other colleagues for their criticisms and corrections. 
\end{acknowledgments}

\section*{DATA AVAILABILITY}
The data sets and codes used in this paper are available by sending request email to the authors at any time.

\section*{CONFLICT OF INTEREST}
All authors declare that they have no conflicts of interest.

\newcommand{\newblock}{}
\bibliographystyle{gji}
\bibliography{srcInvEWI_Ref}

\begin{thebibliography}{32}
\expandafter\ifx\csname natexlab\endcsname\relax\def\natexlab#1{#1}\fi

\bibitem[Abubakar et~al.(2008)Abubakar, Hu, Van Den~Berg, \&
  Habashy]{abubakar2008finite}
Abubakar, A., Hu, W., Van Den~Berg, P., \& Habashy, T., 2008.
\newblock A finite-difference contrast source inversion method, {\it Inverse
  Problems\/}, {\bf 24}(6), 065004.

\bibitem[Alewine~III(1974)]{alewine1974application}
Alewine~III, R.~W., 1974.
\newblock {\it Application of linear inversion theory toward the estimation of
  seismic source parameters\/}, Ph.D. thesis, California Institute of
  Technology.

\bibitem[Artman et~al.(2010)Artman, Podladtchikov, \& Witten]{artman2010source}
Artman, B., Podladtchikov, I., \& Witten, B., 2010.
\newblock Source location using time-reverse imaging, {\it Geophysical
  Prospecting\/}, {\bf 58}(5), 861--873.

\bibitem[Baysal et~al.(1983)Baysal, Kosloff, \& Sherwood]{baysal1983reverse}
Baysal, E., Kosloff, D.~D., \& Sherwood, J.~W., 1983.
\newblock Reverse time migration, {\it Geophysics\/}, {\bf 48}(11), 1514--1524.

\bibitem[Bormann et~al.(2013)Bormann, Wendt, \& DiGiacomo]{bormann2013seismic}
Bormann, P., Wendt, S., \& DiGiacomo, D., 2013.
\newblock Seismic sources and source parameters, in {\em New Manual of
  Seismological Observatory Practice 2 (NMSOP2)\/}, pp. 1--259, Deutsches
  GeoForschungsZentrum GFZ.

\bibitem[Dai et~al.(2011)Dai, Wang, \& Schuster]{dai2011least}
Dai, W., Wang, X., \& Schuster, G.~T., 2011.
\newblock Least-squares migration of multisource data with a deblurring filter,
  {\it Geophysics\/}, {\bf 76}(5), R135--R146.

\bibitem[Fink(2006)]{fink2006time}
Fink, M., 2006.
\newblock Time-reversal acoustics in complex environments, {\it Geophysics\/},
  {\bf 71}(4), SI151--SI164.

\bibitem[Fukahata et~al.(2003)Fukahata, Yagi, \&
  Matsu'ura]{fukahata2003waveform}
Fukahata, Y., Yagi, Y., \& Matsu'ura, M., 2003.
\newblock Waveform inversion for seismic source processes using abic with two
  sorts of prior constraints: Comparison between proper and improper
  formulations, {\it Geophysical research letters\/}, {\bf 30}(6).

\bibitem[Gajewski \& Tessmer(2005)]{gajewski2005reverse}
Gajewski, D. \& Tessmer, E., 2005.
\newblock Reverse modelling for seismic event characterization, {\it
  Geophysical Journal International\/}, {\bf 163}(1), 276--284.

\bibitem[Huang et~al.(2018)Huang, Nammour, \& Symes]{huang2018source}
Huang, G., Nammour, R., \& Symes, W.~W., 2018.
\newblock Source-independent extended waveform inversion based on space-time
  source extension: Frequency-domain implementation, {\it Geophysics\/}, {\bf
  83}(5), R449--R461.

\bibitem[Lamuraglia et~al.(2023)Lamuraglia, Stucchi, \&
  Aleardi]{lamuraglia2023application}
Lamuraglia, S., Stucchi, E., \& Aleardi, M., 2023.
\newblock Application of a global--local full-waveform inversion of rayleigh
  wave to estimate the near-surface shear wave velocity model, {\it Near
  Surface Geophysics\/}, {\bf 21}(1), 21--38.

\bibitem[Lee \& Kim(2003)]{lee2003source}
Lee, K.~H. \& Kim, H.~J., 2003.
\newblock Source-independent full-waveform inversion of seismic data, {\it
  Geophysics\/}, {\bf 68}(6), 2010--2015.

\bibitem[Liu et~al.(2021)Liu, Du, \& Luo]{liu2021sparsity}
Liu, Y., Du, Y., \& Luo, Y., 2021.
\newblock Sparsity-promoting least-squares interferometric migration for
  high-resolution passive source location, {\it Geophysics\/}, {\bf 86}(1),
  KS1--KS9.

\bibitem[Nocedal \& Wright(2006)]{nocedal2006}
Nocedal, J. \& Wright, S., 2006.
\newblock {\it Numerical optimization\/}, Springer Science \& Business Media.

\bibitem[Schuster(2017)]{schuster2017seismic}
Schuster, G.~T., 2017.
\newblock {\it Seismic inversion\/}, Society of Exploration Geophysicists.

\bibitem[Sheng et~al.(2006)Sheng, Leeds, Buddensiek, \&
  Schuster]{sheng2006early}
Sheng, J., Leeds, A., Buddensiek, M., \& Schuster, G.~T., 2006.
\newblock Early arrival waveform tomography on near-surface refraction data,
  {\it Geophysics\/}, {\bf 71}(4), U47--U57.

\bibitem[Sun et~al.(2014)Sun, Jiao, Vigh, \& Coates]{sun2014source}
Sun, D., Jiao, K., Vigh, D., \& Coates, R., 2014.
\newblock Source wavelet estimation in full waveform inversion, in {\em 76th
  EAGE Conference and Exhibition 2014\/}, vol. 2014, pp. 1--5, EAGE
  Publications BV.

\bibitem[Sun et~al.(2016)Sun, Xue, Zhu, Fomel, \& Nakata]{sun2016full}
Sun, J., Xue, Z., Zhu, T., Fomel, S., \& Nakata, N., 2016.
\newblock Full-waveform inversion of passive seismic data for sources and
  velocities, in {\em 2016 SEG International Exposition and Annual Meeting\/},
  OnePetro.

\bibitem[Sun et~al.(2021)Sun, Innanen, \& Huang]{sun2021physics}
Sun, J., Innanen, K.~A., \& Huang, C., 2021.
\newblock Physics-guided deep learning for seismic inversion with hybrid
  training and uncertainty analysis, {\it Geophysics\/}, {\bf 86}(3),
  R303--R317.

\bibitem[Symes(2022)]{symes2022error}
Symes, W.~W., 2022.
\newblock Error bounds for extended source inversion applied to an acoustic
  transmission inverse problem, {\it Inverse Problems\/}, {\bf 38}(11), 115002.

\bibitem[Symes et~al.(2020)Symes, Chen, \& Minkoff]{symes2020full}
Symes, W.~W., Chen, H., \& Minkoff, S.~E., 2020.
\newblock Full-waveform inversion by source extension: Why it works, in {\em
  SEG Technical Program Expanded Abstracts 2020\/}, pp. 765--769, Society of
  Exploration Geophysicists.

\bibitem[Tarantola(1984)]{tarantola1984inversion}
Tarantola, A., 1984.
\newblock Inversion of seismic reflection data in the acoustic approximation,
  {\it Geophysics\/}, {\bf 49}(8), 1259--1266.

\bibitem[Van Den~Berg \& Kleinman(1997)]{van1997contrast}
Van Den~Berg, P.~M. \& Kleinman, R.~E., 1997.
\newblock A contrast source inversion method, {\it Inverse problems\/}, {\bf
  13}(6), 1607.

\bibitem[Van~Leeuwen \& Herrmann(2013)]{van2013mitigating}
Van~Leeuwen, T. \& Herrmann, F.~J., 2013.
\newblock Mitigating local minima in full-waveform inversion by expanding the
  search space, {\it Geophysical Journal International\/}, {\bf 195}(1),
  661--667.

\bibitem[Virieux \& Operto(2009)]{virieux2009overview}
Virieux, J. \& Operto, S., 2009.
\newblock An overview of full-waveform inversion in exploration geophysics,
  {\it Geophysics\/}, {\bf 74}(6), WCC1--WCC26.

\bibitem[Wang et~al.(2009)Wang, Krebs, Hinkley, \&
  Baumstein]{wang2009simultaneous}
Wang, K., Krebs, J.~R., Hinkley, D., \& Baumstein, A., 2009.
\newblock Simultaneous full-waveform inversion for source wavelet and earth
  model, in {\em 2009 SEG Annual Meeting\/}, OnePetro.

\bibitem[Yagi \& Fukahata(2011)]{yagi2011introduction}
Yagi, Y. \& Fukahata, Y., 2011.
\newblock Introduction of uncertainty of green's function into waveform
  inversion for seismic source processes, {\it Geophysical Journal
  International\/}, {\bf 186}(2), 711--720.

\bibitem[Yilmaz(2001)]{yilmaz2001seismic}
Yilmaz, {\"O}., 2001.
\newblock {\it Seismic data analysis: Processing, inversion, and interpretation
  of seismic data\/}, Society of exploration geophysicists.

\bibitem[Yu et~al.(2017)Yu, Zhang, \& Huang]{yu2017application}
Yu, H., Zhang, D., \& Huang, Y., 2017.
\newblock Application of early arrival waveform inversion with
  pseudo-deconvolution misfit function by source convolution, {\it Inverse
  Problems in Science and Engineering\/}, {\bf 25}(1), 57--72.

\bibitem[Zhong \& Liu(2019)]{zhong2019source}
Zhong, Y. \& Liu, Y., 2019.
\newblock Source-independent time-domain vector-acoustic full-waveform
  inversion, {\it Geophysics\/}, {\bf 84}(4), R489--R505.

\bibitem[Zhu \& Ben-Zion(2013)]{zhu2013parametrization}
Zhu, L. \& Ben-Zion, Y., 2013.
\newblock Parametrization of general seismic potency and moment tensors for
  source inversion of seismic waveform data, {\it Geophysical Journal
  International\/}, {\bf 194}(2), 839--843.

\bibitem[Zhu(2014)]{zhu2014time}
Zhu, T., 2014.
\newblock Time-reverse modelling of acoustic wave propagation in attenuating
  media, {\it Geophysical Journal International\/}, {\bf 197}(1), 483--494.

\end{thebibliography}

\clearpage
\clearpage
\clearpage

%
%
%
%
\label{lastpage}

\end{document}